\documentclass[12pt]{article}
\usepackage{amsfonts}
\usepackage{amsmath,amssymb}
\newtheorem{thm}{Theorem}[section]
\newtheorem{prop}[thm]{Proposition}
\newtheorem{lem}[thm]{Lemma}
\numberwithin{equation}{section}
\def\Tr{{\rm Tr\,}}

\def\H{{\cal H}}

\def\Det{{\rm Det}}
\def\R{\Bbb R}
\def\C{\Bbb C}
\def\Z{\Bbb Z}
\def\N{\Bbb N}
\def\SN{{\cal S}_N}
\def\sgn{{\rm sgn}}
\begin{document}

\title {
    A Canonical Ensemble Approach \\
    to the Fermion/Boson Random Point Processes   \\
        and its Applications
             }
\author {   H. Tamura\thanks{tamurah@kenroku.kanazawa-u.ac.jp} \\
 Department of Mathematics, Kanazawa University,\\
          Kanazawa 920-1192, Japan \\
     K. R. Ito\thanks{ito@mpg.setsunan.ac.jp, ito@kurims.kyoto-u.ac.jp} \\
      Department of Mathematics and Physics, Setsunan University,\\
         Neyagawa, Osaka 572-8508, Japan
}
\maketitle
\begin{abstract}
We introduce the  boson and the fermion point 
processes from the elementary quantum mechanical point of view.
That is, we consider quantum statistical mechanics of canonical
ensemble for a fixed number of particles which obey Bose-Einstein, 
Fermi-Dirac statistics, respectively, in a finite volume.
Focusing on the distribution of positions of the particles, we
have point processes of the fixed number of points in a bounded domain.
By taking the thermodynamic limit such that the particle density
converges to a finite value, the boson/fermion processes are obtained.
This argument is a realization of the equivalence of ensembles,
since resulting processes are considered to describe a grand canonical
 ensemble of points.
Random point processes corresponding to para-particles of 
order two are discussed as an application of the formulation.
A statistics of a system of composite particles at zero temperature
are also considered as a model of determinantal random point processes.
\end{abstract}
\section{Introduction}
As special classes of random point processes, fermion point processes 
and boson point processes have been studied by many authors since \cite{BM73, M75, M77}.
Among them, \cite{FF87, F91} made a correspondence between boson processes and 
locally normal states on $C^*$-algebra of operators on the boson Fock 
space.
A functional integral method is used in \cite{L02} to obtain these processes 
from quantum field theories of finite temperatures.
On the other hand, \cite{ST03} formulated both the fermion and boson processes 
in a unified way in terms of the Laplace transformation and generalized them.
Let $Q(R)$ be the space of all the locally finite configurations 
over a Polish space $R$ and $K$ a locally trace class integral operator on  
$L^{2}(R)$ with a Radon measure $\lambda$ on $R$ .
For any nonnegative function $f$ having bounded support and 
$\xi=\sum _{j} \delta_{x_{j}} \in Q(R)$, we set
$<\xi,f>=\sum_{j} f(x_{j})$.
Shirai and Takahashi \cite{ST03} have formulated and studied
the random processes $\mu_{\alpha,K}$ which have Laplace transformations
\begin{equation}
E[e^{-<f,\xi>}]
    \equiv \int_{Q(R)} \mu_{\alpha,K}(d\xi)\,e^{-<\xi,f>}
    =\Det\big(I+\alpha \sqrt{1-e^{-f}}K\sqrt{1-e^{-f}}\big)
                        ^{-1/\alpha}
\label{ft}
\end{equation}
for the parameters $\alpha \in \{2/m;m \in {\Bbb N}\} \cup \{-1/m; m\in N\}$.

\medskip

\noindent Here the cases $\alpha = \pm 1$ correspond to
 boson/fermion processes, respectively.

In their argument, the generalized Vere-Jones' formula\cite{VJ88}
\begin{equation}
    \Det(1-\alpha J)^{-1/\alpha}=\sum \frac{1}{n!}\int_{R^n} 
                  \det{}_{\alpha}(J(x_{i} ,x_{j}))_{i,j=1}^{n}
                  \lambda^{\otimes n}
                  (dx_{1}\cdots dx_{n})
\label{VJ}
\end{equation}
has played an essential role.
Here $J$ is a trace class integral operator, for which we need 
the condition $||\alpha J|| <1$ unless $-1/\alpha \in \N $, 
$ \Det(\,\cdot\,)$ the Fredholm determinant
 and $\det_{\alpha} A$ the $\alpha$-{\it determinant}  defined by
\begin{equation}
\det{}_{\alpha}A=\sum_{\sigma\in {\cal S}_{n}}\alpha^{n-\nu(\sigma)}
                              \prod_{i} A_{i\sigma(i)}
\label{adet}
\end{equation}
for  a matrix  $A$ of size $n\times n$, where 
$\nu(\sigma)$ is the numbers of cycles in $\sigma$.
The formula (\ref{VJ}) is Fredholm's original definition of his functional 
determinant in the case $\alpha = -1$.

The purpose of the paper is to construct both the fermion and boson
processes from a view point of elementary quantum mechanics in order to
get simple, clear and straightforward understanding of them in the 
 connection with physics. 
Let us consider the system of $N$ free fermions/bosons in a 
box of finite volume $V$ in $\R^d$ and the quantum statistical mechanical
state of the system with a finite temperature.
Giving the distribution function of the positions of all particles in terms of 
the square of the absolute value of the wave functions, 
we obtain a point process of $N$ points in the box.
As the thermodynamic limit, $N, V \to \infty$ and $N/V \to \rho$, of these 
processes of finite points, fermion and boson processes in $\R^d$
with density $\rho$ are obtained.
In the argument, we will use the generalized Vere-Jones' formula in the form:
\begin{equation}
      \frac{1}{N!}\int 
                  \det{}_{\alpha}(J(x_{i} ,x_{j}))_{i,j=1}^{N}
                  \lambda^{\otimes N}
                  (dx_{1}\cdots dx_{N})
      = \oint _{S_r(0)}\frac{dz}{2\pi iz^{N+1}}\Det(1- z\alpha J)^{-1/\alpha},
\label{IVJ}
\end{equation}
where $r>0$ is arbitrary for $-1/\alpha \in \N$, otherwise $r$ should satisfy 
$||r\alpha J||<1$.
Here and hereafter, $S_r(\zeta)$ denotes the integration contour defined
by the map  $ \theta \mapsto \zeta + r\exp(i\theta) $, 
where $\theta$ ranges from $-\pi$ to $\pi$, $ r>0 $ and $ \zeta \in \C$.
In the terminology of statistical mechanics, we start from canonical ensemble 
and end up with formulae like (\ref{ft}) and (\ref{VJ}) of grand canonical nature.
In this sense, the argument is related to the equivalence of ensembles.
The use of (\ref{IVJ}) makes our approach simple.  

In this approach, we need neither quantum field theories nor the theory 
of states on the operator algebras to derive the boson/fermion processes.
It is interesting to apply the method to the problems which have not
been formulated in statistical mechanics on quantum field theories yet.
Here, we study the system of para-fermions and para-bosons of order 2.
Para statistics was first introduced by Green\cite{G53} in the context of 
quantum field theories.
For its review, see \cite{OK82}.
\cite{MG64} and \cite{HT69, ST70} formulated it within the framework of 
quantum mechanics of finite number of particles.  See also \cite{OK69}.
Recently statistical mechanics of para-particles are formulated
in \cite{S90, C96, CS97}. However, it does not seem to be fully 
developed so far.
We formulate here point processes as the distributions of positions of 
para-particles of order 2 with finite temperature and positive density 
through the thermodynamic limit.
It turns out that the resulting processes are corresponding to the 
cases $\alpha = \pm 1/2$ in \cite{ST03}. 
We also try to derive point processes from ensembles of composite particles at
zero temperature and positive density in this formalism.
The resulting processes also have their Laplace transforms expressed by 
Fredholm determinants.

This paper is organized as follows.
In Section 2, the random point processes of fixed numbers of fermions
as well as bosons are formulated on the base of quantum mechanics in a bounded box.
Then, the theorems on thermodynamic limits are stated.
The proofs of the theorems are presented in Section 3 as applications
of a theorem of rather abstract form.
In Sections 4 and 5, we consider the systems of para-particles and composite
particles, respectively.
In Appendix, we calculate complex integrals needed for the thermodynamic
limits.

\section{Fermion and boson processes} 

Consider $ L^2(\Lambda_L) $ on $ \Lambda_L = [-L/2, L/2]^d $ 
$\subset {\Bbb R}^d$ with the Lebesgue measure on $\Lambda_L$.  
Let $\triangle_L$ be the Laplacian  on $ {\cal H}_L = L^2(\Lambda_L) $ 
satisfying periodic boundary conditions at $\partial \Lambda_{L}$.  
We deal with periodic boundary conditions in this paper, however, 
all the arguments except that in section 5 may be applied
for other boundary conditions.
Hereafter we regard $ -\triangle_L$ as  the quantum mechanical 
Hamiltonian of a single free particle.
The usual factor $\hbar^2/2m$ is set at unity.
For $k\in {\Bbb Z}^d$, 
$ \varphi_k^{(L)}(x) = L^{-d/2} \exp (i2\pi k\cdot x/L) $ 
is an eigenfunction of $\triangle_L$,  and 
$ \, \{\, \varphi_k^{(L)} \, \}_{k\in {\Bbb Z}^d} \,$ forms an 
complete orthonormal system [CONS] of $ \H_L $.
In the following, we use the operator $G_L = \exp(\beta\triangle_L)$ 
whose kernel is given by
\begin{equation}
     G_L(x,y) = \sum_{k\in {\Bbb Z}^d}e^{-\beta  |2\pi k/L|^2} \varphi_k^{(L)}(x)
     \overline{\varphi_k^{(L)}(y)} 
\label{GL}
\end{equation}
for $\beta > 0$.  We put $ g_k^{(L)} = \exp (-\beta |2\pi k/L|^2) $, the 
eigenvalue of $G_L$ for the eigenfunction $\varphi_k^{(L)}$.
We also need $G = \exp(\beta \triangle)$ on $L^2({\Bbb R}^d)$ and its kernel
\[
    G(x,y) = \int_{ {\Bbb R}^d}\frac{dp}{(2\pi)^d}
    e^{-\beta |p|^2 +ip\cdot(x-y)} 
           = \frac{\exp(-|x-y|^2/4\beta)}{(4\pi \beta)^{d/2}}.
\]
Note that $G_L(x,y)$ and $G(x,y)$ are real symmetric and 
\begin{equation}
     G_L(x,y) = \sum_{k\in {\Bbb Z}^d} G(x, y+kL).
\label{GG}
\end {equation}

Let $ f: {\Bbb R}^d \rightarrow [0,\infty) $ be an arbitrary continuous function 
whose support is compact.
In the course of the thermodynamic limit,
$f$ is fixed and we assume that $L$ is so large that $\Lambda_L $
contains the support, and regard $f$ as a function on $\Lambda_L$.

\subsection{Fermion processes}
In this subsection, we construct the fermion process in $\R^d$ as a limit of
the process of $N$ points in $\Lambda_L$.
Suppose there are $N$ identical particles which obey the Fermi-Dirac statistics
in a finite box $\Lambda_L$.
The space of the quantum mechanical states of the system is given by
\[
 {\cal H}^F_{L,N} = \{ \, A_Nf \,| \, f \in \otimes^N{\cal H}_L\, \},  
\]
where 
\[
      A_N f(x_1, \cdots, x_N) = \frac{1}{N!}\sum_{\sigma \in \SN} 
      \sgn(\sigma) f(x_{\sigma(1)}, \cdots, x_{\sigma(N)})
      \qquad (\; x_1, \cdots, x_N \in \Lambda_L \;)
\]
is anti-symmetrization in the $N$ indices.
Using the CONS $  \{\, \varphi_k^{(L)} \, \}_{k\in {\Bbb Z}^d} \,$
of $\H_L = L^2(\Lambda_L)$, we make the element 
\begin{equation}
   \Phi_k(x_1, \cdots, x_N) 
    = \frac{1}{\sqrt{N!}}\sum_{\sigma\in\SN} \sgn(\sigma)
      \varphi_{k_1}(x_{\sigma(1)})\cdot \cdots \cdot 
      \varphi_{k_N}(x_{\sigma(N)})
\label{fcons}
\end{equation}
of ${\cal H}^F_{L,N}$ for $ k=(k_1, \cdots, k_N) \in (\Z^d)^N$.
Let us introduce the lexicographic order $\prec$ in $\Z^d$ and put
$ (\Z^d)^N_{\precneqq} = \{ \, (k_1, \cdots, k_N) \in (\Z^d)^N \,|
  \, k_1 \precneqq \cdots \precneqq k_N \, \}$. 
Then $ \{\,\Phi_k \,\}_{k\in (\Z^d)^N_{\precneqq}} $ 
forms a CONS of ${\cal H}^F_{L,N}$.

According to the idea of the canonical ensemble in quantum statistical
mechanics, the probability density distribution of the positions of 
the $N$ free fermions in the periodic box $\Lambda_L$ 
at the inverse temperature $\beta$ is given by
\begin{eqnarray}
 p^F_{L, N}(x_1, \cdots, x_N) 
    &=& Z_F^{-1}\sum_{k\in (\Z^d)^N_{\precneqq}} 
      \Big(\prod_{j=1}^N g_{k_j}^{(L)}\Big) 
      |\Phi_k(x_1, \cdots, x_N)|^2     \nonumber \\
   &=& Z_F^{-1}\sum_{k\in (\Z^d)^N_{\precneqq}} 
        \overline{\Phi_k(x_1, \cdots, x_N)}\big((\otimes^NG_L)\Phi_k\big)
        (x_1, \cdots, x_N)
\label{ffp}
\end{eqnarray}
where $Z_F$ is the normalization constant.
We can define the point process of $N$ points in $\Lambda_L$ from
the density (\ref{ffp}).  I.e., consider a map 
$ \Lambda_L^N \ni ( x_1, \cdots, x_N) \mapsto 
             \sum_{j=1}^N \delta_{x_j} \in Q(\R^d) $.
Let $ \mu_{L, N}^F $ be the probability measure on $Q(\R^d)$ 
induced by the map from the probability measure on $\Lambda_L^N$ 
which has the density (\ref{ffp}).
By $ {\rm E}^F_{L,N} $, we denote expectation with respect to the measure 
$ \mu_{L, N}^F $.
The Laplace transform of the point process is given by
\begin{eqnarray}
     {\rm E}^F_{L,N}\big[e^{-<f,\xi>}\big]
     &=&\int_{Q(\R^d)}d\mu_{L, N}^F(\xi)\,e^{-<f,\xi>}
                          \nonumber \\
     &=& \int_{\Lambda_L^N}\exp(-\sum_{j=1}^Nf(x_j))
       p_{L, N}^F(x_1, \cdots, x_N) \,dx_1\cdots dx_N \nonumber \\
     &=&  \frac{\Tr_{\H_{L,N}^F}[(\otimes^Ne^{-f})(\otimes^NG_L)]}
          {\Tr_{\H_{L,N}^F}[\otimes^N G_L]}
                          \nonumber \\
     &=& \frac{\Tr_{\otimes^N\H_L}[(\otimes^N\tilde G_L)A_N]}
          {\Tr_{\otimes^N\H_L}[(\otimes^N G_L)A_N]}
\label{fgfl}   \\
     &=& \frac{\int_{\Lambda_L^N}
       \det_{-1}\tilde G_L(x_i, x_j)\,dx_1\cdots dx_N}
     {\int_{\Lambda_L^N}\det_{-1}G_L(x_i, x_j)\,dx_1\cdots dx_N},
         \nonumber 
\end{eqnarray}
where $\tilde G_L$ is defined by
\begin{equation}
       \tilde G_L = G_L^{1/2}e^{-f}G_L^{1/2},
\label{tildeg}
\end{equation}
where
$e^{-f}$ represents the operator of multiplication by the function $e^{-f}$.

The fifth expression follows from 
$ [ \otimes^NG_L^{1/2}, A_N] = 0 $, cyclicity of the trace and 
 $(\otimes^NG_L^{1/2})$
$(\otimes^Ne^{-f})(\otimes^NG_L^{1/2}) 
= \otimes^N\tilde G_L$ and so on.
The last expression can be obtained by calculating the trace 
on $\otimes^N \H_L$ using its CONS 
     $\{\varphi_{k_1}\otimes\cdots \otimes\varphi_{k_N} \, 
     | \, k_1, \cdots, k_N \in \Z^d \}$, 
where $\det_{-1}$ is the  usual determinant, see eq. (\ref{adet}).

Now, let us consider the thermodynamic limit, where the volume 
of the box $\Lambda_L$ and the number of points $N$ in the box 
$\Lambda_L$ tend to infinity in such a way that the densities tend
 to a positive finite value $\rho $, i.e.,
\begin{equation}
       L,\; N \rightarrow \infty, \quad N/L^d \to \rho > 0.
\label{tdl}
\end{equation}

\begin{thm}
The finite fermion processes $\{ \, \mu_{L,N}^F \, \}$ defined above
 converge weakly to the fermion process  $\mu_{\rho}^F$ whose Laplace 
transform is given by 
\begin{equation}
      \int_{Q(\R^d)}e^{-<f, \xi>}d\mu_{\rho}^F(\xi)
       = \Det \big[1 - \sqrt{1-e^{-f}}z_*G(1+z_*G)^{-1}\sqrt{1-e^{-f}}\big]
\label{EF}
\end{equation}
in the thermodynamic limit (\ref{tdl}), where $z_*$ is the positive number
uniquely determined by
\[
  \rho = \int \frac{dp}{(2\pi)^d}\frac{z_*e^{-\beta |p|^2}}
  {1 + z_*e^{-\beta |p|^2}}
  =(z_*G(1+z_*G)^{-1})( x, x).
\]
\label{fthm}
\end{thm}
{\sl Remark : } The existence of $\mu_{\rho}^F$ which has the 
above Laplace transform is a consequence of the result of \cite{ST03}
 we have mentioned in the introduction. 
\subsection{Boson processes}
Suppose there are $N$ identical particles which obey Bose-Einstein
statistics in a finite box $\Lambda_L$.
The space of the quantum mechanical states of the system is given by
\[
 {\cal H}^B_{L,N} = \{ \, S_Nf \,| \, f \in \otimes^N{\cal H}_L\, \},
\]
where
\[
      S_N f(x_1, \cdots, x_N) = \frac{1}{N!}\sum_{\sigma \in \SN}
       f(x_{\sigma(1)}, \cdots x_{\sigma(N)})
      \qquad ( \; x_1, \cdots, x_N \in \Lambda_L \; )
\]
is symmetrization in the $N$ indices.
Using the CONS $  \{\, \varphi_k^{(L)} \, \}_{k\in {\Bbb Z}^d} \,$
of $L^2(\Lambda_L)$, we make the element
\begin{equation}
      \Psi_k(x_1, \cdots, x_N) = \frac{1}{\sqrt{N!n(k)}}\sum_{\sigma\in\SN}
      \varphi_{k_1}(x_{\sigma(1)})\cdot \cdots \cdot
      \varphi_{k_N}(x_{\sigma(N)})
\label{bcons}
\end{equation}
of ${\cal H}^B_{L,N}$ for $ k=(k_1, \cdots, k_N) \in \Z^d$, 
where $ n(k) = \prod_{l\in\Z^d}(\sharp\{\,n\in\{ \, 1, \cdots, N\,\} \,|\,
k_n = l
\,\}!)$.
Let us introduce the subset
 $ (\Z^d)^N_{\prec} = \{ \, (k_1, \cdots, k_N) \in (\Z^d)^N \,|
     \, k_1 \prec \cdots \prec k_N \, \} $ of $(\Z^d)^N$.  
Then $ \{\,\Psi_k \,\}_{k\in (\Z^d)^N_{\prec}} $ forms a CONS of
${\cal H}^B_{L,N}$.

As in the fermion's case,
the probability density distribution of the positions of the $N$ free
bosons in the periodic box $\Lambda_L$ at the inverse temperature 
$\beta$ is given by
\begin{equation}
      p^B_{L, N}(x_1, \cdots, x_N)
      = Z_B^{-1}\sum_{k\in (\Z^d)^N_{\prec}}
       \Big(\prod_{j=1}^N g_{k_j}^{(L)}\Big)
      |\Psi_k(x_1, \cdots, x_N)|^2,
\label{d+1}
\end{equation}
where $Z_B$ is the normalization constant.

We can define a point process of $N$ points $\mu_{L,N}^B $ 
from the density (\ref{d+1}) as in the previous section.
The Laplace transform of the point process is given by
\begin{equation}
     {\rm E}^B_{L,N}\big[e^{-<f,\xi>}\big]
     = \frac{\Tr_{\otimes^N{\cal H}_L}[(\otimes^N\tilde G_L)S_N]}
      {\Tr_{\otimes^N{\cal H}_L}[(\otimes^NG_L)S_N]}
     = \frac{\int_{\Lambda_L^N}\det_{1}\tilde G_L(x_i, x_j)\,dx_1\cdots
      dx_N}
     {\int_{\Lambda_L^N}\det_{1}G_L(x_i, x_j)\,dx_1\cdots dx_N},
\label{bgfl}
\end{equation}
where det$_1$ denotes permanent, see eq. (\ref{adet}). We set 
\begin{equation}
      \rho_c = \int_{\R^d} \frac{dp}{(2\pi)^d}\frac{e^{-\beta |p|^2}}
         {1 - e^{-\beta |p|^2}},
\label{rho_c}
\end{equation}
which is finite for $d > 2$.
Now, we have
\begin{thm}
The finite boson processes $\{ \, \mu_{L,N}^B \, \}$  defined above converge
 weakly to the boson process $\mu_{\rho}^B $ whose Laplace transform is given by
\begin{equation}
    \int_{Q(\R^d)} e^{-<f, \xi>} d\mu_{\rho}^B(\xi)
       = \Det [1 + \sqrt{1-e^{-f}}z_*G(1 - z_*G)^{-1}\sqrt{1-e^{-f}}]^{-1}
\label{EB}
\end{equation}
in the thermodynamic limit (\ref{tdl}) if
\[
      \rho = \int_{\R^d} \frac{dp}{(2\pi)^d}\frac{z_*e^{-\beta |p|^2}}
    {1 - z_*e^{-\beta |p|^2}} = (z_*G(1 - z_*G)^{-1})(x,x) < \rho_c.
\]
\label{bthm}
\end{thm}

\medskip

\noindent {\sl Remark 1 : } For the existence of $\mu_{\rho}^B$ , we 
refer to \cite{ST03}.

\noindent {\sl Remark 2 : } In this paper, we only consider the boson processes
with low densities : $\rho < \rho_c$.  
The high density cases $\rho \geqslant \rho_c$ are related to the
Bose-Einstein condensation.  We need the detailed knowledge about
the spectrum of $\tilde G_L$ to deal with these cases.
It will be reported in another publication.

\section{ Thermodynamic limits}
\subsection{A general framework}
It is convenient to consider the problem in a general framework on a  Hilbert
space $\H$ over $ \C$.  
The proofs of the theorems of section 2 are given in the next subsection.
We denote the operator norm by $||\,\cdot\,||$, the trace norm
by $||\,\cdot\,||_1$ and the Hilbert-Schmidt norm by $||\,\cdot\,||_2$.
Let $\{ V_L\}_{L > 0}$ be a one-parameter family of Hilbert-Schmidt
operators on $\H$ which satisfy the conditions
\[
        \forall L>0: || \, V_L \, ||=1, \quad
        \lim_{L\to\infty}|| \, V_L \, ||_2=\infty
\]
and  $A$ a bounded self-adjoint operator on $\H$ satisfying
$ 0 \leqslant A \leqslant 1 $.
Then $G_L = V_L^*V_L, \; \tilde G_L= V_L^*AV_L$ are
self-adjoint trace class operators satisfying
\[
    \forall L>0 : 0 \leqslant \tilde G_L \leqslant G_L \leqslant 1, \; ||G_L||=1
       \; \mbox{ and } \; \lim_{L\to\infty}\Tr G_L = \infty.
\]
We define $ I_{-1/n}= [0, \infty)$ for $ n\in \N$
and $ I_{\alpha}=[ 0, 1/|\alpha|)$ for
$\alpha \in [ -1, 1]-\{ 0, -1, -1/2, \cdots \}$.
Then the function
\[
     h_L^{(\alpha)}(z)=\frac{\Tr[zG_L(1-z\alpha G_L)^{-1}]}{\Tr G_L}
\]
is well defined on $ I_{\alpha}$ for each $L>0$ and $\alpha\in [ -1, 1] -\{0\}$.

\medskip

\begin{thm}
Let $\alpha \in [ -1, 1]-\{0\}$ be arbitrary but fixed.
Suppose that for every $z\in I_{\alpha}$, there exist a limit
$h^{(\alpha)}(z)=\lim_{L\to\infty}h_L^{(\alpha)}(z)$ and a trace class operator
$K_z$ satisfying
\begin{equation}
     \lim_{L\to\infty}|| \, K_z - (1-A)^{1/2}V_L(1-z\alpha
            V_L^*V_L)^{-1}V_L^*(1-A)^{1/2}||_1 = 0.
\label{K_z}
\end{equation}
Then, for every $ \hat\rho\in [0, \sup_{z\in I_{\alpha}}h^{(\alpha)}(z))$, there exists
a unique solution $z=z_*\in I_{\alpha}$ of $ h^{(\alpha)}(z) =\hat\rho$.
Moreover suppose that a sequence $L_1 < L_2 < \cdots < L_N < \cdots $ 
satisfies
\begin{equation}
\lim_{N\to\infty}N/\Tr G_{L_N} = \hat\rho.
\label{gtdl}
\end{equation}
Then
\begin{equation}
   \lim_{N\to\infty}\frac{\sum_{\sigma\in\SN}\alpha^{N-\nu(\sigma)}
         \Tr_{\otimes^N\H}[\otimes^N\tilde G_{L_N} U(\sigma)]}
        {\sum_{\sigma\in\SN}\alpha^{N-\nu(\sigma)}
         \Tr_{\otimes^N\H}[\otimes^N G_{L_N} U(\sigma)]}
        = \Det[1+z_*\alpha K_{z_*}]^{-1/\alpha}
\label{limDet}
\end{equation}
holds.
\label{gthm}
\end{thm}

\medskip
\noindent In (\ref{limDet}), the operator $U(\sigma)$ on $\otimes^N \H$ is
defined by
$ U(\sigma)\varphi_1 \otimes \cdots \otimes \varphi_N =
  \varphi_{\sigma^{-1}(1)} \otimes \cdots \otimes \varphi_{\sigma^{-1}(N)}$
for $\sigma \in\SN$ and $ \varphi_1, \cdots, \varphi_N \in \H$.
In order to prove the theorem, we prepare several lemmas
under the same assumptions of the theorem.
\begin{lem}
$h^{(\alpha)}$ is a strictly increasing continuous function on $I_{\alpha}$ and
there exists a unique $z_* \in I_{\alpha}$ which satisfies 
$ h^{(\alpha)}(z_*) =\hat\rho$.
\label{P1}
\end{lem}
{\sl Proof : } From ${h^{(\alpha)}_L}'(z)=\Tr [G_L(1-z\alpha G_L)^{-2}]/\Tr G_L$, 
we have
$1 \leqslant {h_L^{(\alpha)}}'(z)\leqslant (1-z\alpha)^{-2}$ for $\alpha > 0 $
and $1 \geqslant {h_L^{(\alpha)}}'(z)\geqslant (1-z\alpha)^{-2}$ for $\alpha < 0 $,
i.e., $\{h_L^{(\alpha)}\}_{\{L>0\}}$ is equi-continuous on $I_{\alpha}$.
By Ascoli-Arzel\`a's theorem, the convergence
 $h_L^{(\alpha)} \to h^{(\alpha)} $ is locally uniform and hence $h^{(\alpha)}$ is 
 continuous on $I_{\alpha}$.
It also follows that $h^{(\alpha)}$ is strictly increasing.
Together with $h^{(\alpha)}(0)=0$, which comes from $h_L^{(\alpha)}(0)=0$, we get that
 $h^{(\alpha)}(z)=\hat\rho$ has a unique solution in $I_{\alpha}$.  
\hfill $\Box$

\begin{lem}
There exists a constant $c_0 > 0$ such that
\[
    ||G_L - \tilde G_L||_1 = \Tr [V_L^*(1-A)V_L] \leqslant c_0
\]
uniformly in $L > 0$.
\label{P2}
\end{lem}
{\sl Proof : } Since \; $ 1-z\alpha G_L $ \; is invertible for $ z \in
I_{\alpha}$
and $V_L$ is Hilbert-Schmidt,
we have
\begin{eqnarray}
\lefteqn{\Tr [V_L^*(1-A)V_L] } && \nonumber \\
&=& \Tr [(1-z\alpha G_L)^{1/2} (1-z\alpha G_L)^{-1/2}
 V_L^*(1-A)V_L(1-z\alpha G_L)^{-1/2} (1-z\alpha G_L)^{1/2}]
                                \nonumber  \\
&\leqslant & ||1-z\alpha G_L|| \Tr[(1-z\alpha G_L)^{-1/2}V_L^*(1-A)V_L
     (1-z\alpha G_L)^{-1/2}]    \nonumber \\
&=& ||1-z\alpha G_L|| \Tr[(1-A)^{1/2}V_L (1-z\alpha
                 G_L)^{-1}V_L^*(1-A)^{1/2}] \nonumber \\
&=& (1-(\alpha\wedge 0)z)(\Tr K_z +o(1)).
\end{eqnarray}
Here we have used $|\Tr B_1CB_2| \leqslant ||B_1||\,||B_2||\,||C||_1 
=||B_1||\,||B_2||\Tr C$ for bounded
 operators $B_1, B_2$ and a positive trace class operator $C$
and $ \Tr WV = \Tr VW $ for Hilbert-Schmidt operators $W, V$.
\hfill  $\Box$

Let us denote all the eigenvalues of  $G_L$ and $\tilde G_L$ in 
decreasing order
\[
        g_0(L) =1 \geqslant  g_1(L) \geqslant \cdots \geqslant
         g_j(L) \geqslant \cdots
\]
and
\[
        \tilde g_0(L) \geqslant \tilde g_1(L) \geqslant \cdots \geqslant
        \tilde g_j(L) \geqslant \cdots,
\]
respectively.
Then we have
\begin{lem}
For each $ j = 0, 1, 2, \cdots, \quad
  g_j(L) \geqslant \tilde g_j(L) $ \quad holds.
\label{minmax}
\end{lem}

\noindent{\sl Proof:}  By the min-max principle, we have
\begin{eqnarray*}
\tilde g_j(L) &=& \min_{\psi_0, \cdots, \psi_{j-1} \in {\cal H}_L}
      \; \max_{\psi \in \{\psi_0, \cdots, \psi_{j-1}\}^{\perp}}
      \frac{(\psi, \tilde G_L\psi)}{||\psi||^2} \nonumber \\
      &\leqslant & 
      \min_{\psi_0, \cdots, \psi_{j-1} \in {\cal H}_L}
      \; \max_{\psi \in \{\psi_0, \cdots, \psi_{j-1}\}^{\perp}}
      \frac{(\psi, G_L \psi)}{||\psi||^2} = g_j(L).
      \mbox{\hspace*{5cm}}    \Box
\end{eqnarray*}

\begin{lem}
For $N$ large enough, the conditions
\begin{equation}
   \Tr [z_NG_{L_N}(1-\alpha z_NG_{L_N})^{-1}]=
   \Tr [\tilde z_N\tilde G_{L_N}(1-\alpha \tilde z_N\tilde G_{L_N})^{-1}]=N
\label{n}
\end{equation}
determine $z_N, \tilde z_N \in I_{\alpha}$ uniquely.
$z_N$ and $ \tilde z_N $ satisfy
\[
    z_N \leqslant \tilde z_N, \quad | \tilde z_N -z_N |=O(1/N) \quad
    \mbox{and} \quad \lim_{N\to\infty}z_N=\lim_{N\to\infty}\tilde z_N=z_*.
\]
\label{P4}
\end{lem}
{\sl Proof : } From the proof of Lemma \ref{P1},
 $H_N(z) = \Tr [zG_{L_N}(1-z\alpha G_{L_N})^{-1}] = h_{L_N}^{(\alpha)}(z)\Tr G_{L_N}$
is a strictly increasing continuous function on $I_{\alpha}$ and $H_N(0)=0$.
Let us pick $z_0 \in I_{\alpha}$ such that $z_0 > z_*$.
Since $h^{(\alpha)}$ is strictly increasing, $ h^{(\alpha)}(z_0) - h^{(\alpha)}(z_*) = \epsilon > 0$.
We have
\begin{equation}
     \frac{H_N(z_0)}{N}=\frac{\Tr G_{L_N}}{N}h_{L_N}(z_0) \to
     \frac{h^{(\alpha)}(z_0)}{\hat\rho} = 1 + \frac{\epsilon}{\hat\rho},
\label{h_N}
\end{equation}
which shows $H_N\big((\sup I_{\alpha})-0\big) \geqslant H_N(z_0) > N$ for large enough
$N$.
Thus $z_N \in[0, z_0) \subset I_{\alpha}$ is uniquely determined by 
$H_N(z_N) = N$.

Put  $\tilde H_N(z) = \Tr [z\tilde G_{L_N}(1-z\alpha \tilde G_{L_N})^{-1}]
$. 
Then  by Lemma \ref{minmax}, $\tilde H_N$ is well-defined on $I_{\alpha}$
and
$\tilde H_N \leqslant H_N$ there.
Moreover
\begin{eqnarray*}
H_N(z) -\tilde H_N(z)
       &=& \Tr [(1-\alpha zG_{L_N})^{-1}
         z(G_{L_N}-\tilde G_{L_N})(1-\alpha z\tilde G_{L_N})^{-1}] \\
       &\leqslant& ||(1-\alpha zG_{L_N})^{-1}||
        ||(1-\alpha z\tilde G_{L_N})^{-1}||z\Tr [G_{L_N}- \tilde G_{L_N}]\\
       &\leqslant & C_z = \frac{zc_0}{(1-(\alpha\vee 0) z)^2}
\end{eqnarray*}
holds.
Together with (\ref{h_N}), we have
\[
       \frac{\tilde H_N(z_0)}{N} \geqslant \frac{ H_N(z_0) - C_{z_0}}{N} >
        1+ \frac{\epsilon}{2\hat \rho} - \frac{C_{z_0}}{N},
\]
hence $\tilde H_N(z_0) > N$, if $N$ is large enough.
It is also obvious that $\tilde H_N$ is  strictly increasing and 
continuous on $I_{\alpha}$ and $\tilde H_N(0)=0$.
Thus $\tilde z_N \in [0, z_0) \subset I_{\alpha}$ is uniquely determined by
 $ \tilde H_N(\tilde z_N)=N$.

The convergence $ z_N \to z_* $ is a consequence of
$h_{L_N}^{(\alpha)}(z_N) = N/\Tr G_{L_N} \to \hat \rho =h^{(\alpha)}(z_*)$,
the strict increasingness of $h^{(\alpha)}, h_L^{(\alpha)}$ and the pointwise
convergence  $ h_L^{(\alpha)} \to h^{(\alpha)}$.
We get $ z_N \leqslant\tilde z_N$ from $ H_N \geqslant \tilde H_N $ and the
increasingness of $H_N, \tilde H_N$.

Now, let us show $ |\tilde z_N-z_N|=O(N^{-1})$, which together 
with $z_N \to z_*$, yields $\tilde z_N \to z_*$.
From
\begin{eqnarray*}
0  &=& N - N = H_N(z_N) - \tilde H_N(\tilde z_N) \\
   &=& \Tr[(1-\alpha z_NG_{L_N})^{-1}(z_NG_{L_N}
     - \tilde z_N\tilde G_{L_N})(1-\alpha \tilde z_N\tilde G_{L_N})^{-1}]\\
   &=& z_N\Tr[(1-\alpha z_NG_{L_N})^{-1}(G_{L_N} - \tilde G_{L_N})
     (1-\alpha \tilde z_N\tilde G_{L_N})^{-1}] \\
   & & 
      -(\tilde z_N -z_N)\Tr[(1-\alpha z_NG_{L_N})^{-1}\tilde G_{L_N}
      (1-\alpha \tilde z_N\tilde G_{L_N})^{-1}],
\end{eqnarray*}
we get
\begin{eqnarray*}
\lefteqn
{\frac{\tilde z_N - z_N}{\tilde z_N}
   \Tr[(1-\alpha z_NG_{L_N})^{-1/2}\tilde z_N\tilde G_{L_N}
   (1-\alpha \tilde z_N\tilde G_{L_N})^{-1}(1-\alpha z_NG_{L_N})^{-1/2}]
 }  && \\
&=&z_N\Tr[(1-\alpha z_NG_{L_N})^{-1}(G_{L_N}-\tilde G_{L_N})
      (1-\alpha \tilde z_N\tilde G_{L_N})^{-1}].
\end{eqnarray*}
It follows that
\begin{eqnarray*}
  \lefteqn{\frac{\tilde z_N - z_N}{\tilde z_N}N =
    \frac{\tilde z_N - z_N}{\tilde z_N}\tilde H_N(\tilde z_N) }&&\\
&=&
    \frac{\tilde z_N - z_N}{\tilde z_N}
   \Tr[(1-\alpha z_NG_{L_N})(1-\alpha z_NG_{L_N})^{-1/2}\tilde z_N\tilde
G_{L_N}
      (1-\alpha \tilde z_N\tilde G_{L_N})^{-1}(1-\alpha z_NG_{L_N})^{-1/2}]
\end{eqnarray*}
\begin{eqnarray*}
   & \leqslant & \frac{\tilde z_N - z_N}{\tilde z_N} ||1-\alpha z_NG_{L_N}||
   \Tr[(1-\alpha z_NG_{L_N})^{-1/2}\tilde z_N\tilde G_{L_N}
      (1-\alpha \tilde z_N\tilde G_{L_N})^{-1}(1-\alpha z_NG_{L_N})^{-1/2}]
\\
    & = & ||1-\alpha z_NG_{L_N}|| z_N\Tr[(1-\alpha z_NG_{L_N})^{-1}
        (G_{L_N}-\tilde G_{L_N})(1-\alpha \tilde z_N\tilde G_{L_N})^{-1}]
\\
    & \leqslant & z_N ||1-\alpha z_NG_{L_N}|| \, ||(1-\alpha z_NG_{L_N})^{-1}||\,
    \Tr [G_{L_N}-\tilde G_{L_N}]\,||(1-\alpha \tilde z_N\tilde
G_{L_N})^{-1}||
\\
    & \leqslant & c_0z_0(1-(\alpha\wedge 0) z_0)/(1-(\alpha\vee 0)z_0)^2
\end{eqnarray*}
for $N$ large enough, because $z_N, \tilde z_N <z_0$.
Thus, we have obtained $\,\tilde z_N -z_N = O(N^{-1})$. \hfill  $\Box$

\bigskip

We put
\[
      v^{(N)} =  \Tr[z_NG_{L_N}(1-\alpha z_NG_{L_N})^{-2}] \quad \mbox{and}
\quad
           \tilde v^{(N)} =  \Tr[\tilde z_N\tilde G_{L_N}(1-\alpha \tilde
z_N
              \tilde G_{L_N})^{-2}].
\]
Then we have :
\begin{lem}
${\rm (i)}  \quad  \displaystyle v^{(N)}, \tilde v^{(N)} \to \infty,$
   \hspace*{3cm}
 ${\rm (ii)} \quad  \displaystyle
       \frac{v^{(N)}}{\tilde v^{(N)}} \to 1.$
\label{vv}
\end{lem}
{\sl Proof : } (i) follows from the lower bound
\begin{eqnarray*}
     v^{(N)} &=&  \Tr[z_NG_{L_N}(1-\alpha z_NG_{L_N})^{-2}]
\\
     &\geqslant&  \Tr[z_NG_{L_N}(1-\alpha z_NG_{L_N})^{-1}]\,
                   ||1-\alpha z_NG_{L_N}||^{-1}
     \geqslant  N(1+o(1))/(1-(\alpha\wedge 0) z_*),
\end{eqnarray*}
since $z_N\to z_*$. The same bound is also true for $\tilde v^{(N)}$.

\noindent 
(ii) Using
\begin{eqnarray*}
      v^{(N)} &=& \Tr[-z_NG_{L_N}(1-\alpha z_NG_{L_N})^{-1}
               +\alpha^{-1}(1-\alpha z_NG_{L_N})^{-2} - \alpha^{-1}]
\\
         &=& -N + \alpha^{-1}\Tr[(1-\alpha z_NG_{L_N})^{-2} - 1]
\end{eqnarray*}
and the same for $\tilde v^{(N)}$, we get
\begin{eqnarray*}
|\tilde v^{(N)} - v^{(N)}|
     &=&  |\alpha^{-1}\Tr[(1 -\alpha \tilde z_N\tilde G_{L_N})^{-2}
        - (1-\alpha z_NG_{L_N})^{-2} ]| \\
     &\leqslant&  |\alpha^{-1}\Tr[\big((1 -\alpha \tilde z_N\tilde G_{L_N})^{-1}
    - (1 -\alpha z_N\tilde G_{L_N})^{-1}\big)
    (1 -\alpha  \tilde z_N\tilde G_{L_N})^{-1}]|\\
     & &        + |\alpha^{-1}\Tr[\big((1 -\alpha z_N\tilde G_{L_N})^{-1} -
        (1 -\alpha z_N G_{L_N})^{-1}\big)
        (1 -\alpha  \tilde z_N\tilde G_{L_N})^{-1}]| \\
     & &     + |\alpha^{-1}\Tr[(1 -\alpha z_N G_{L_N})^{-1}
        \big((1 -\alpha \tilde z_N\tilde G_{L_N})^{-1} -
        (1 -\alpha z_N\tilde G_{L_N})^{-1}\big)]| \\
     & &    + |\alpha^{-1}\Tr[(1 -\alpha z_N G_{L_N})^{-1}
        \big((1 -\alpha z_N\tilde G_{L_N})^{-1} -
        (1 -\alpha z_N G_{L_N})^{-1}\big)]| \\
     &\leqslant&  ||(1 -\alpha \tilde z_N\tilde G_{L_N})^{-1}||
       \, ||(1 -\alpha \tilde z_N\tilde G_{L_N})^{-1}
             (\tilde z_N -z_N)\tilde G_{L_N}
        (1 -\alpha z_N\tilde G_{L_N})^{-1}||_1 \\
      & & + ||(1 -\alpha \tilde z_N\tilde G_{L_N})^{-1}|| \,
        ||(1 -\alpha z_N\tilde G_{L_N})^{-1}
        z_N(G_{L_N} - \tilde G_{L_N})(1 -\alpha z_N G_{L_N})^{-1}||_1 \\
      & &  + ||(1 -\alpha  z_N G_{L_N})^{-1}|| \,
        ||(1 -\alpha \tilde z_N\tilde G_{L_N})^{-1}
        (\tilde z_N -z_N)\tilde G_{L_N}
          (1 -\alpha  z_N\tilde G_{L_N})^{-1}||_1 \\
       & &  + ||(1 -\alpha z_N G_{L_N})^{-1}|| \,
        ||(1 -\alpha z_N\tilde G_{L_N})^{-1}
        z_N(G_{L_N} - \tilde G_{L_N})(1 -\alpha z_N G_{L_N})^{-1}||_1 \\
       & \leqslant& ( ||(1 -\alpha \tilde z_N\tilde G_{L_N})^{-1}||
     + ||(1 -\alpha z_N G_{L_N})^{-1}||) \\
       & &       \times\bigg(\frac{\tilde z_N - z_N}{\tilde z_N}
          ||\tilde z_N\tilde G_{L_N}(1-\alpha \tilde z_N\tilde
         G_{L_N})^{-1}||_1 
          \, ||(1 -\alpha  z_N\tilde G_{L_N})^{-1}|| \\
       & &   + z_N||(1 -\alpha z_N\tilde G_{L_N})^{-1}||\,
     ||G_{L_N} - \tilde G_{L_N}||_1
     \,|| (1 - \alpha z_N G_{L_N})^{-1}||\bigg) = O(1).
\end{eqnarray*}
In the last step, we have used Lemmas \ref{P2} and \ref{P4}.
This, together with (i), implies (ii).  \hfill   $\Box$

\begin{lem}
\[
      \lim_{N\to \infty}\sqrt{2\pi v^{(N)}}\oint_{S_1(0)}
      \frac{d\eta}{2\pi i\eta^{N+1}}\Det\big[1 -\alpha z_N (\eta-1) G_{L_N}
      (1-\alpha z_NG_{L_N})^{-1}\big]^{-1/\alpha}  = 1,
\]
\[
      \lim_{N\to \infty}\sqrt{2\pi \tilde v^{(N)}}\oint_{S_1(0)}
     \frac{d\eta}{2\pi i\eta^{N+1}}
         \Det\big[1 -\alpha \tilde z_N (\eta-1) \tilde G_{L_N}
      (1-\alpha \tilde z_N\tilde G_{L_N})^{-1}\big]^{-1/\alpha}  = 1,
\]
\label{intDet}
\end{lem}
{\sl Proof : } Put $ s = 1/|\alpha| $ and
\[
    p_j^{(N)} = \frac{|\alpha|z_Ng_j(L_N)}{1 -\alpha  z_Ng_j(L_N)}.
\]
Then the first equality is nothing but proposition A.2(i) for
$\alpha < 0$ and proposition A.2(ii) for $\alpha >0$.
The same is true for the second equality.
\hfill $\Box$

\bigskip
{\sl Proof of Theorem \ref{gthm} : }
Since the uniqueness of $z_*$ has already been shown, it is enough to
prove (\ref{limDet}).
The main apparatus of the proof is Vere-Jones' formula in 
the following form:  Let $\alpha = -1/n$ for $n \in \N$. 
Then
\[
    \Det(1-\alpha J)^{-1/\alpha} = \sum_{n=0}^{\infty}\frac{1}{n!}
   \sum_{\sigma\in{\cal S}_n}\alpha^{n-\nu(\sigma)}
         \Tr_{\otimes^n\H}[(\otimes^n J) U(\sigma)]
\]
holds for any trace class operator $J$.
For $\alpha \in[ -1, 1]- \{ 0,  -1, -1/2, \cdots, 1/n, \cdots\} $,
this holds under an additional condition $||\alpha J|| <1$.
This has actually been proved in Theorem 2.4 of \cite{ST03}.
We use the formula in the form
\begin{equation}
   \frac{1}{N!} \sum_{\sigma\in\SN}\alpha^{N-\nu(\sigma)}
         \Tr_{\otimes^N\H}[(\otimes^NG_{L_N}) U(\sigma)]
    = \oint_{S_{z_N}(0)}\frac{dz}{2\pi iz^{N+1}}
    \Det(1-z\alpha G_{L_N})^{-1/\alpha}
\label{gIVJ}
\end{equation}
and in the form in which $G_{L_N}$ is replaced by $\tilde G_{L_N}$.
Here, recall that $z_N, \tilde z_N \in I_{\alpha}$.
We calculate the right-hand side by the saddle point method.

Using the above integral representation and the property of
the  products of the  Fredholm determinants followed by the change 
of integral variables $z = z_N\eta, z= \tilde z_N\eta$, we get
\[
         \frac{\sum_{\sigma\in\SN}\alpha^{N-\nu(\sigma)}
         \Tr_{\otimes^N\H}[\otimes^N \tilde G_{L_N} U(\sigma)]}
        {\sum_{\sigma\in\SN}\alpha^{N-\nu(\sigma)}
         \Tr_{\otimes^N\H}[\otimes^N G_{L_N} U(\sigma)]}
        = \frac{\oint_{S_{\tilde z_N}(0)}
         \Det(1-z\alpha \tilde G_{L_N})^{-1/\alpha}dz/2\pi iz^{N+1}}
         {\oint_{S_{z_N}(0)}
         \Det(1- z\alpha G_{L_N})^{-1/\alpha}dz/2\pi iz^{N+1}}
\]
\begin{eqnarray*}
& =&  \frac{\Det[1- \tilde z_N\alpha G_{L_N}]^{-1/\alpha}}
              {\Det[1- z_N\alpha G_{L_N}]^{-1/\alpha}}
         \frac{\Det[1- \tilde z_N\alpha \tilde G_{L_N}]^{-1/\alpha}}
              {\Det[1- \tilde z_N\alpha G_{L_N}]^{-1/\alpha}}
         \frac{z_N^N}{\tilde z_N^N} \\
& &      \times \frac{\oint_{S_1(0)}
         \Det[1-\tilde z_N(\eta-1)\alpha \tilde G_{L_N}
          (1-\tilde z_N\alpha \tilde G_{L_N})^{-1}]^{-1/\alpha}
         d\eta/2\pi i\eta^{N+1}}
         {\oint_{S_1(0)}
         \Det[1-z_N(\eta-1)\alpha G_{L_N}
           (1-z_N\alpha G_{L_N})^{-1}]^{-1/\alpha}d\eta/2\pi i\eta^{N+1}}.
\end{eqnarray*}
Thus the theorem is proved if the following behaviors 
in $N \to \infty$ are valid: 
\begin{eqnarray*}
{\rm (a)}& \mbox{\hspace{0.5cm}}
              &\frac{z_N^N}{\tilde z_N^N} \; = \; 
           \exp\big(- \frac{\tilde z_N - z_N}{z_N}N +o(1)\big) 
\\
{\rm (b)}&          & \frac{\Det[1-\tilde z_N\alpha G_{L_N}]^{-1/\alpha}}
              {\Det[1-  z_N\alpha G_{L_N}]^{-1/\alpha}} \; = \;
               \exp\big(\frac{\tilde z_N - z_N}{z_N}N +o(1)\big),
\\
{\rm (c)}&          &\frac{\Det[1-\tilde z_N\alpha \tilde G_{L_N}]^{-1/\alpha}}
              {\Det[1- \tilde z_N\alpha G_{L_N}]^{-1/\alpha}} \; \to \; 
                  \Det[1+z_*\alpha K_{z_*}]^{-1/\alpha}
\\
{\rm (d)}&          &\frac{\oint_{S_1(0)}\Det[1-\tilde z_N(\eta-1)\alpha 
                \tilde G_{L_N}(1-\tilde z_N\alpha \tilde G_{L_N})^{-1}]^{-1/\alpha}
                d\eta/2\pi i\eta^{N+1}}
         {\oint_{S_1(0)}\Det[1-z_N(\eta-1)\alpha G_{L_N}
           (1-z_N\alpha G_{L_N})^{-1}]^{-1/\alpha}d\eta/2\pi i\eta^{N+1}}
     \quad \to \; 1.
\end{eqnarray*}
In fact, (a) is a consequence of Lemma \ref{P4}. 
For (b), let us define a function 
  $ k(z) = \log\,\Det[1-z\alpha G_L]^{-1/\alpha} = 
  -\alpha^{-1}\sum_{j=0}^{\infty}\log(1-z\alpha g_j(L))$.
Then by Taylor's formula and (\ref{n}), we get
\[
  k(\tilde z_N) - k(z_N) = k'(z_N)(\tilde z_N -z_N) + k''(\bar z)
  \frac{( \tilde z_N - z_N)^2}{2} \mbox{\hspace{3cm}}
\]
\[
  = \sum_{j=0}^{\infty}\frac{g_j}{1- z_N\alpha g_j}(\tilde z_N - z_N)
  + \sum_{j=0}^{\infty}\frac{\alpha g_j^2}{(1-\bar z\alpha g_j)^2}
  \frac{( \tilde z_N - z_N)^2}{2}
  = N\frac{ \tilde z_N - z_N}{z_N} + \delta,
\]
where $ \bar z $ is a mean value of $z_N$ and $\tilde z_N$ and 
$ |\delta| = O(1/N) $ by Lemma \ref{P4}.

From the property of the product and the cyclic nature of the
Fredholm determinants, we have 
\begin{eqnarray*}
\lefteqn{ \frac{\Det[1-\tilde z_N\alpha \tilde G_{L_N}]}
              {\Det[1- \tilde z_N\alpha G_{L_N}]}} && \\
  &=&  \Det[1+ z_*\alpha (1-A)^{1/2}V_{L_N}
         (1- z_*\alpha G_{L_N})^{-1}V_{L_N}^*(1-A)^{1/2}] \\
 & &  + \big\{ \Det[1+ \tilde z_N\alpha(G_{L_N} - \tilde G_{L_N})
              (1- \tilde z_N\alpha G_{L_N})^{-1}] \\
 & & 
             - \Det[1+  z_*\alpha(G_{L_N} - \tilde G_{L_N})
              (1- z_*\alpha G_{L_N})^{-1}] \big\}.
\end{eqnarray*}
The first term converges to $\Det[1+z_*\alpha K_{z_*}]$ by
the assumption (\ref{K_z}) and the continuity of the
Fredholm determinants with respect to the trace norm.
The brace in the above equation tends to $0$, because of the continuity
and 
\begin{eqnarray*}
\lefteqn{||\tilde z_N\alpha(G_{L_N} - \tilde G_{L_N})
              (1- \tilde z_N\alpha G_{L_N})^{-1}
             - z_*\alpha(G_{L_N} - \tilde G_{L_N})
              (1- z_*\alpha G_{L_N})^{-1}||_1 } && \\
& & \leqslant |\tilde z_N-z_*|\,|\alpha|\,||G_{L_N} - \tilde G_{L_N}||_1
              ||(1- \tilde z_N\alpha G_{L_N})^{-1}|| \\
& &  +  z_*|\alpha|\,||G_{L_N} - \tilde G_{L_N}||_1
 ||(1- \tilde z_N\alpha G_{L_N})^{-1} - (1- z_*\alpha G_{L_N})^{-1}||
 \to 0, 
\end{eqnarray*}
where we have used Lemmas \ref{P2} and \ref{P4}.
Thus, we get (c).
(d) is a consequence of Lemma \ref{vv} and Lemma \ref{intDet}.
\hfill  

\subsection{Proofs of the theorems}
To prove Theorem \ref{fthm}[\ref{bthm}], it is enough to show that
(\ref{fgfl})[(\ref{bgfl})] converges to the right-hand side of (\ref{EF})
[(\ref{EB}), respectively] for every $ f\in C_o(\R^d)$\cite{DVJ}.
We regard $ \H_L = L^2(\Lambda_L) $ as a closed subspace of $L^2(\R^d)$.
Corresponding to the orthogonal decomposition 
$ L^2(\R^d) = L^2(\Lambda_L) \oplus L^2(\Lambda_L^c) $, we set 
$V_L = e^{\beta\triangle_L/2} \oplus 0$.
Let $ A = e^{-f} $ be the multiplication operator on $L^2(\R^d)$,
which can be decomposed as 
$ A = e^{-f}\chi_{\Lambda_L} \oplus \chi_{\Lambda_L^c}$
for large $L$ since supp$\,f$ is compact.
Then
\[
      G_L = V_L^*V_L = e^{\beta\triangle_L} \oplus 0 \quad\mbox{and}\quad
      \tilde G_L = V_L^*AV_L = e^{\beta\triangle_L/2}e^{-f}
         e^{\beta\triangle_L/2} \oplus 0
\]
can be identified with those in section 2.

We begin with the following fact, where we denote
\[
            \Box_k^{(L)} \; = \; \frac{2\pi}{L}\Big(k + 
          \Big( -\frac{1}{2}, \frac{1}{2}\Big]^d\Big)
         \qquad \mbox{for} \quad k \in \Z^d. 
\]
\begin{lem}
Let $ b : [0, \infty) \to [0, \infty) $ be a monotone decreasing 
continuous function such that
\[
       \int_{\R^d}b(|p|)dp < \infty.
\]
Define the function $ b_L : \R^d \to [0, \infty)$ by
\[
        b_L(p) =
          b(|2\pi k/L|) \qquad \mbox{ if } \quad
     p\in \Box_k^{(L)} 
     \quad \mbox{for} \quad k \in \Z^d. 
\]
Then $ b_L(p) \to b(|p|) $ in $L^1(\R^d)$ as $L \to \infty $ .
\end{lem}
{\sl Proof : } There exist positive constants $c_1$ and $c_2$ such that
$ b_L(p) \leqslant c_1b(c_2|p|) $ holds for all $L\geqslant 1$ and $ p \in \R^d$.
Indeed, $c_1 = b(0)/b(2\pi\sqrt{d/(d+8)}), \, c_2 = 2/\sqrt{d+8}$ satisfy the condition,
 since $ \inf\{ \, c_1b(c_2|p|) \, | \, p \in \Box_0^{(L)} \, \} \geqslant b(0) $
for $\forall L >1$ and $ \sup\{ \,c_2|p| \, | \, p \in \Box_k^{(L)} \, \} 
\leqslant 2\pi|k|/L $ for $ k \in \Z^d -\{0\}$.
Obviously $ c_1b(c_2|p|) $ is an integrable function of $ p \in \R^d$.
The lemma follows by the dominated convergence theorem. \hfill    $\Box$

\medskip

Finally we confirm the assumptions of theorem \ref{gthm}.
\begin{prop}
\begin{eqnarray}
{\rm (i)} &&   \forall L>0:  ||V_L|| = 1, \qquad
  \lim_{L\to\infty} \Tr G_L/L^d =(4\pi \beta)^{-d/2}.
               \label{TrG_L} \\
{\rm (ii)}&& \mbox{The following convergences hold as } L\to \infty 
\mbox{ for each } z\in I_{\alpha}:\hskip2cm
                             \notag\\
&& \mbox{\hspace{-1cm}} 
   h_L^{(\alpha)}(z) = 
 \frac{\Tr [zG_L(1-z\alpha G_L)^{-1}]}{\Tr G_L}
 \; \to \; (4\pi \beta)^{d/2}\int_{\R^d}\frac{dp}{(2\pi)^d}
       \frac{ze^{-\beta|p|^2}}{1-z\alpha e^{-\beta|p|^2}} = h^{(\alpha)}(z),
       \label{falpha} \\
&& \mbox{\hspace{-1cm}}
  ||\sqrt{1-e^{-f}}\big( G_L(1-z\alpha G_L)^{-1}
   - G(1-z\alpha G)^{-1}\big)\sqrt{1-e^{-f}}||_1\to 0.
\label{limK}
\end{eqnarray}
\end{prop}
{\sl Proof : }
By applying the above lemma to $  b(|p|) = e^{-\beta|p|^2} $ and 
$ \tilde b(|p|) = ze^{-\beta|p|^2}/(1-z\alpha e^{-\beta|p|^2})$, 
we have (\ref{TrG_L}) and (\ref{falpha}). 

By Gr\"um's convergence theorem, it is enough to show 
$$ \sqrt{1-e^{-f}} G_L(1-z\alpha G_L)^{-1}\sqrt{1-e^{-f}}
\to \sqrt{1-e^{-f}} G(1-z\alpha G)^{-1}\sqrt{1-e^{-f}} $$
strongly and
\begin{eqnarray*}
\lefteqn{ \Tr [\sqrt{1-e^{-f}}G_L(1-z\alpha G_L)^{-1}\sqrt{1-e^{-f}}] = 
     \int_{\R^d}(1-e^{-f(x)}) \big(G_L(1-z\alpha G_L)^{-1}\big)(x, x)dx
        }  && \\
&\to& \int_{\R^d}(1-e^{-f(x)}) \big(G(1-z\alpha G)^{-1}\big)(x, x)dx
     = \Tr[\sqrt{1-e^{-f}} G(1-z\alpha G)^{-1}\sqrt{1-e^{-f}}]
\end{eqnarray*}
for (\ref{limK}). These are direct consequences of
\begin{eqnarray*}
\lefteqn{| zG_L(1-z\alpha G_L)^{-1}(x,y)  -  zG(1-z\alpha G)^{-1}(x,y) | } &&  \\ 
&=& \int\frac{dp}{(2\pi)^d}|e_L(p, x-y)
             \tilde b_L(p) -e(p, x-y)\tilde b(|p|)| \\
&\leqslant& \int\frac{dp}{(2\pi)^d}\big(|\tilde b_L(p)-\tilde b(|p|)| + 
      |e_L(p, x-y) - e(p, x-y)|\tilde b(|p|)\big) \to 0
\end{eqnarray*}
uniformly in $x ,y \in$ supp $f$.
Here we have used the above lemma for $\tilde b(|p|)$ and we put
 $e(p,x) = e^{ip\cdot x}$ and 
\begin{equation}
      e_L(p; x) = e(2\pi k/L; x) \quad\mbox{if} \quad  
     p\in \Box_k^{(L)}
     \quad \mbox{for} \quad k \in \Z^d.
\tag*{$\Box$}
\end{equation}
Thanks to (\ref{TrG_L}), we can take a sequence $\{ L_N\}_{N\in \N}$
which satisfies (\ref{gtdl}).
On the relation between $\rho$ in Theorems \ref{fthm}, \ref{bthm} 
and $\hat\rho$ in Theorem \ref{gthm}, $\hat\rho=(4\pi\beta)^{d/2}\rho$
 is derived from (\ref{tdl}).
We have the ranges of $\rho$ in Theorem \ref{bthm} and Theorem \ref{fthm}, since
\[
       \sup_{z\in I_1}h^{(1)}(z)= (4\pi\beta)^{d/2}\int_{\R^d}\frac{dp}{(2\pi)^d}
       \frac{e^{-\beta|p|^2}}{1- e^{-\beta|p|^2}}
       = (4\pi\beta)^{d/2}\rho_c
\]
and $ \sup_{z\in I_{-1}}h^{(-1)}(z)=\infty$ from (\ref{falpha}).
Thus we get Theorem \ref{fthm} and Theorem \ref{bthm} using Theorem \ref{gthm}.

\section{Para-particles}
The purpose of this section is to apply the method which we have developed
in the preceding sections to statistical mechanics of gases which consist of 
identical particles obeying para-statistics.
Here, we restrict our attention to para-fermions and para-bosons of
order $2$.
We will see that the point processes obtained after the thermodynamic limit
are the point processes corresponding to the cases of $ \alpha = \pm 1/2 $
given in \cite{ST03}.

In this section, we use the representation theory of the symmetric
group ( cf. e.g. \cite{JK, Sa91, Si96}). 
We say that $ (\lambda_1, \lambda_2, \cdots, \lambda_n) \in {\Bbb N}^n $ is
a Young frame of length $n$ for the symmetric group ${\cal S}_N$ if
\[
   \sum_{j=1}^n\lambda_j =N,  \quad
    \lambda_1 \geqslant \lambda_2 \geqslant  \cdots \geqslant \lambda_n >0.
\]
We associate the Young frame $ (\lambda_1, \lambda_2, \cdots, \lambda_n) $ with
the diagram of $\lambda_1$-boxes in the first row, $\lambda_2$-boxes in the
second row,..., and $\lambda_n$-boxes in the $n$-th row.
A Young tableau on a Young frame is a bijection from the numbers $1, 2,
\cdots, N$ to the $N$ boxes of the frame.

\subsection{Para-bosons of order 2}

Let us select one Young tableau, arbitrary but fixed, on each Young 
frame of length less than or equal to 2, say the tableau $T_j$ on the 
frame $( N-j, j)$ for $ j = 1, 2, \cdots, [N/2]$ and the tableau 
$T_0$ on the frame $(N)$.
We denote by ${\cal R}(T_j)$ the row stabilizer of $T_j$, i.e., 
the subgroup of ${\cal S}_N$ consists of those elements that keep all 
rows of $T_j$ invariant, and by ${\cal C}(T_j)$ the column stabilizer 
whose elements preserve all
columns of $T_j$.

Let us introduce the three elements
\[
   a(T_j)=\frac{1}{\#{\cal R}(T_j)}\sum_{\sigma \in {\cal R}(T_j)}\sigma,
\qquad
   b(T_j)=\frac{1}{\#{\cal C}(T_j)}\sum_{\sigma \in {\cal C}(T_j)}
   \sgn(\sigma)\sigma
\]
and
\[
   e(T_j)= \frac{d_{T_j}}{N !}\sum_{\sigma \in {\cal R}(T_j)}
   \sum_{\tau \in {\cal C}(T_j)}\sgn(\tau)\sigma\tau
        = c_ja(T_j)b(T_j)
\]
of the group algebra 
${\Bbb C}[{\cal S}_N]$ for each $j=0, 1, \cdots, [N/2]$,
where $d_{T_j}$ is the dimension of the irreducible representation of
${\cal S}_N$ corresponding to $T_j$ and 
$ c_j = d_{T_j}\#{\cal R}(T_j)\#{\cal C}(T_j)/N ! $.
As is known,  
\begin{equation}
a(T_j)\sigma b(T_k)=b(T_k)\sigma a(T_j) = 0
\label{asb}
\end{equation}
hold for any $\sigma \in\SN$ and $0\leqslant j<k\leqslant [N/2]$.
The relations
\begin{equation}
   a(T_j)^2 = a(T_j), \quad b(T_j)^2 =b(T_j), \quad
        e(T_j)e(T_k)=\delta_{jk}e(T_j)
\label{abe}
\end{equation}
also hold.
For later use, let us introduce
\begin{equation}
       d(T_j) = e(T_j)a(T_j) = c_j a(T_j)b(T_j)a(T_j) 
       \qquad ( j=0, 1, \cdots,[N/2]).
\label{defd}
\end{equation}
They satisfy
\begin{equation}
      d(T_j)d(T_k)=\delta_{jk}d(T_j) \quad \mbox{ for } \quad  0\leqslant j, k \leqslant
[N/2],
\label{ddd}
\end{equation}
as is shown readily from (\ref{asb}) and (\ref{abe}).
The inner product $< \cdot, \cdot>$ of $\C[\SN]$ is defined by
\[
    < \sigma, \tau> = \delta_{\sigma\tau} \quad \mbox{ for } \; \sigma, \tau
\in\SN
\]
and extended to all elements of $\C[\SN]$ by sesqui-linearity.

The left representation $L$ and the right representation $R$ of $\SN$ 
on $\C[\SN]$ are defined by
\[
    L(\sigma)g = L(\sigma)\sum_{\tau\in\SN}g(\tau)\tau
             =\sum_{\tau\in\SN}g(\tau)\sigma\tau
             = \sum_{\tau\in\SN}g(\sigma^{-1}\tau)\tau
\]
and
\[
    R(\sigma)g = R(\sigma)\sum_{\tau\in\SN}g(\tau)\tau
             =\sum_{\tau\in\SN}g(\tau)\tau\sigma^{-1}
             = \sum_{\tau\in\SN}g(\tau\sigma)\tau,
\]
respectively. Here and hereafter we identify $g: \SN \to \C$ and
$\sum_{\tau\in\SN}g(\tau)\tau \in\C[\SN]$.
They are extended to the representation of $\C[\SN]$ on $\C[\SN]$ as
\[
   L(f)g = fg =\sum_{\sigma,\tau}f(\sigma)g(\tau)\sigma\tau
         =\sum_{\sigma}\big(\sum_{\tau}f(\sigma\tau^{-1})g(\tau)\big)\sigma
\]
and
\[
   R(f)g = g\hat f =\sum_{\sigma,\tau}g(\sigma)f(\tau)\sigma\tau^{-1}
         =\sum_{\sigma}\big(\sum_{\tau}g(\sigma\tau)f(\tau)\big)\sigma,
\]
where $\hat f = \sum_{\tau}\hat f(\tau)\tau = \sum_{\tau}f(\tau^{-1})\tau
         = \sum_{\tau}f(\tau)\tau^{-1}$.

The character of the irreducible representation of $\SN$ corresponding
to the tableau $T_j$ is obtained by
\[
    \chi_{T_j}(\sigma)=\sum_{\tau\in\SN}(\tau, \sigma R(e(T_j))\tau)
    = \sum_{\tau\in\SN}(\tau, \sigma \tau \widehat{e(T_j)}).
\]
We introduce a tentative notation
\begin{equation}
        \chi_{g}(\sigma) \equiv \sum_{\tau\in\SN}(\tau, \sigma R(g)\tau)
    = \sum_{\tau,\gamma\in\SN}(\tau, \sigma \tau \gamma^{-1})g(\gamma)
    = \sum_{\tau\in\SN}g(\tau^{-1}\sigma\tau)
\label{chig}
\end{equation}
for $ g=\sum_{\tau}g(\tau)\tau\in\C[\SN]$.

Let $U$ be the representation of $\SN$ ( and its extension to
$\C[\SN]$) on $\otimes^N\H_L$ defined by
\[
      U(\sigma) \varphi_1\otimes \cdots \otimes \varphi_N =
    \varphi_{\sigma^{-1}(1)}\otimes \cdots \otimes\varphi_{\sigma^{-1}(N)}
        \qquad \mbox{for } \; \varphi_1, \cdots, \varphi_N \in\H_L,
\]
or equivalently by
\[
       (U(\sigma) f)(x_1, \cdots, x_N ) =
    f(x_{\sigma(1)}, \cdots,x_{\sigma(N)})
        \qquad \mbox{ for } \; f \in \otimes^N\H_L.
\]
Obviously, $U$ is unitary: $ U(\sigma)^* = U(\sigma^{-1}) = U(\sigma)^{-1}$.
Hence $U(a(T_j))$ is an orthogonal projection because of
$ U(a(T_j))^* = U(\widehat{a(T_j)}) = U(a(T_j)) $ and (\ref{abe}).
So are $U(b(T_j))$'s, $U(d(T_j))$'s and $P_{2B} = \sum_{j=0}^{[N/2]}U(d(T_j))$.
Note that Ran$\,U(d(T_j)) = \,$Ran$\,U(e(T_j))$ because of 
$d(T_j)e(T_j) = e(T_j), e(T_j)d(T_j)= d(T_j)$. 

We refer the literatures \cite{MG64, HT69, ST70} for quantum mechanics
of para-particles. 
(See also \cite{OK69}.)
The arguments of these literatures indicate that the state space of 
$N$ para-bosons of order 2 in the finite box $\Lambda_L$ is given by
$\H_{L,N}^{2B} = P_{2B}\otimes^N\H_L$. 
It is obvious that there is a CONS of $\H_{L,N}^{2B}$ which consists of 
the vectors of the form 
$U(d(T_j))\varphi_{k_1}^{(L)}\otimes \cdots \otimes \varphi_{k_N}^{(L)}$, 
which are the eigenfunctions of $\otimes^NG_L$.
Then, we define a point process of $N$ free para-bosons of order 2 as 
in section 2 and its generating functional is given by
\[
   E_{L, N}^{2B}\big[e^{-<f, \xi>}\big] =
   \frac{\Tr_{\otimes^N\H_L}
            [(\otimes^N\tilde G_L) P_{2B}]}
        {\Tr_{\otimes^N\H_L}
         [(\otimes^N G_L)P_{2B}]}.
\]
Let us give expressions, which have a clear correspondence 
with (\ref{bgfl}).
\begin{lem}
  \begin{eqnarray}
  E_{L, N}^{2B}\big[e^{-<f, \xi>}\big]
    &=&
   \frac{\sum_{j=0}^{[N/2]}\sum_{\sigma\in\SN}\chi_{T_j}(\sigma)
         \Tr_{\otimes^N\H_L}[(\otimes^N\tilde G_L) U(\sigma)]}
                {\sum_{j=0}^{[N/2]}\sum_{\sigma\in\SN}\chi_{T_j}(\sigma)
         \Tr_{\otimes^N\H_L}[(\otimes^N G_L) U(\sigma)]}
   \label{pgfl}    \\
   &=&\frac{\sum_{j=0}^{[N/2]}\int_{\Lambda_L^N}
           \det_{T_j}\{\tilde G_L(x_i, x_j)\}dx_1 \cdots dx_N}
           {\sum_{j=0}^{[N/2]}\int_{\Lambda_L^N}
           \det_{T_j}\{ G_L(x_i, x_j)\}dx_1 \cdots dx_N}
   \label{imt}  
   \end{eqnarray}
\label{pl}
\end{lem}
{\sl Remark 1. } $ \H_{L,N}^{2B} = P_{2B}\otimes^N\H_L $ is determined by
the choice of the tableaux $T_j$'s.
The spaces corresponding to different choices of tableaux are different 
subspaces of $\otimes^N\H_L$.
However, they are unitarily equivalent and the generating functional
given above is not affected by the choice.
In fact, $\chi_{T_j}(\sigma)$ depends only on
the frame on which the tableau $T_j$ is defined.

\noindent{\sl Remark 2. }  det$_{T}A = \sum_{\sigma\in\SN}
\chi_T(\sigma)\prod_{i=1}^NA_{i\sigma(i)}$ in (\ref{imt}) is called immanant, 
another generalization of determinant than $\det_{\alpha}$.

\medskip

\noindent {\sl Proof : }  Since $\otimes^N G$ commutes with $U(\sigma)$ and 
$ a(T_j)e(T_j) = e(T_j)$, we have
\[
      \Tr_{\otimes^N\H_L}\big((\otimes^N G_L) U(d(T_j))\big)
       = \Tr_{\otimes^N\H_L}\big((\otimes^N G_L) U(e(T_j))U(a(T_j))\big)
       \mbox{\hspace{2cm}}
\]
\begin{equation}
       = \Tr_{\otimes^N\H_L}
             \big(U(a(T_j))(\otimes^N G_L) U(e(T_j))\big)
       = \Tr_{\otimes^N\H_L}
             \big((\otimes^N G_L) U(e(T_j))\big).
\label{d-e}
\end{equation}
On the other hand, we get from (\ref{chig}) that
\begin{eqnarray}
\lefteqn{
 \sum_{\sigma\in\SN}\chi_g(\sigma)
   \Tr_{\otimes^N\H_L}\big((\otimes^NG)U(\sigma)\big)
   =\sum_{\tau, \sigma\in\SN}g(\tau^{-1}\sigma\tau)
   \Tr_{\otimes^N\H_L}\big((\otimes^NG)U(\sigma)\big)
        } &&  \nonumber \\
&=& \sum_{\tau,\sigma}g(\sigma)
   \Tr_{\otimes^N\H_L}\big((\otimes^NG)U(\tau\sigma\tau^{-1})\big)
   = \sum_{\tau,\sigma}g(\sigma)
   \Tr_{\otimes^N\H_L}\big((\otimes^NG)U(\tau)U(\sigma)U(\tau^{-1})\big)
         \nonumber  \\
&=& N!\sum_{\sigma}g(\sigma)
   \Tr_{\otimes^N\H_L}\big((\otimes^NG)U(\sigma)\big)
   = N!\Tr_{\otimes^N\H_L}\big((\otimes^NG)U(g)\big),
\label{imm}
\end{eqnarray}
where we have used the cyclicity of the trace and the commutativity 
of $U(\tau)$ with $\otimes^NG$.  Putting $g=e(T_j)$ and using (\ref{d-e}), the first equation
is derived. The second one is obvious. \hfill $\Box$

\medskip

Let $\psi_{T_j}$ be the character of the induced representation 
Ind$_{{\cal R}(T_j)}^{\SN}[{\bf 1}] $, where {\bf 1} is the 
representation ${\cal R}(T_j) \ni \sigma \to 1$, i.e.,
\[
         \psi_{T_j}(\sigma) = \sum_{\tau\in\SN}< \tau, \sigma
        R(a(T_j))\tau> = \chi_{a(T_j)}(\sigma).
\]
Then the determinantal form \cite{JK}
\begin{eqnarray}
   \chi_{T_j}  &=& \psi_{T_j} - \psi_{T_{j-1}} \qquad ( j= 1, \cdots, [N/2])
\label{chipsi} \\
    \chi_{T_0}  &=& \psi_{T_0}
\notag
\end{eqnarray}
yields the following result:
\begin{thm}
The finite para-boson processes defined above converge weakly to the
point process whose Laplace transform is given by
\[
      {\rm E}_{\rho}^{2B}\big[e^{-<f, \xi>}\big]
       = \Det \big[1 + \sqrt{1-e^{-f}}z_*G(1 - z_*G)^{-1}\sqrt{1-e^{-f}}\big]^{-2}
\]
in the thermodynamic limit, where $z_* \in (0, 1)$ is determined by
\[
     \frac{\rho}{2} = \int \frac{dp}{(2\pi)^d}\frac{z_*e^{-\beta |p|^2}}
    {1 - z_*e^{-\beta |p|^2}} = (z_*G(1 - z_*G)^{-1})(x,x) < \rho_c,
\]
and $\rho_c$ is given by (\ref{rho_c}).
\end{thm}

{\sl Proof : } Using (\ref{chipsi}) in the  expression in
Lemma \ref{pl} and (\ref{imm}) for $ g = a(T_{[N/2]})$, we have
\begin{eqnarray*}
   E_{L, N}^{2B}\big[e^{-<f, \xi>}\big] 
   &=&
         \frac{\sum_{\sigma\in\SN}\psi_{T_{[N/2]}}(\sigma)
           \Tr_{{\cal H}_L^{\otimes N}}
        \big((\otimes^N\tilde G_L)U(\sigma)\big)}
           {\sum_{\sigma\in\SN}\psi_{T_{[N/2]}}(\sigma)
           \Tr_{{\cal H}_L^{\otimes N}}
        \big((\otimes^N G_L)U(\sigma)\big)} \nonumber \\
  &=& \frac{\Tr_{\otimes^N\H_L}
        \big((\otimes^N\tilde G_L)U(a(T_{[N/2]})\big)}
         {\Tr_{\otimes^N\H_L}
        \big((\otimes^N G_L)U(a(T_{[N/2]})\big)} \nonumber \\
  & =&
      \frac{\Tr_{\otimes^{[(N+1)/2]}\H_L}
        \big((\otimes^{[(N+1)/2]}\tilde G_L)S_{[(N+1)/2]}\big)}
         {\Tr_{\otimes^{[(N+1)/2]}\H_L }
        \big((\otimes^{[(N+1)/2]} G_L)S_{[(N+1)/2]}\big)}
      \frac{\Tr_{\otimes^{[N/2]}\H_L}
        \big((\otimes^{[N/2]}\tilde G_L)S_{[N/2]}\big)}
           {\Tr_{\otimes^{[N/2]}\H_L}
        \big((\otimes^{[N/2]} G_L)S_{[N/2]}\big)}.
\end{eqnarray*}
In the last equality, we have used
\[
     a(T_{[N/2]}) = \frac{\sum_{\sigma\in{\cal R}_1}\sigma}{\#{\cal R}_1}
                    \frac{\sum_{\tau\in{\cal R}_2}\tau}{\#{\cal R}_2},
\]
where $ {\cal R}_1 $ is the symmetric group of $[(N+1)/2]$ numbers which are on 
the first row of the tableau $T_{[N/2]}$ and $ {\cal R}_2 $ that of $[N/2]$ 
numbers on the second row.  
Then, Theorem \ref{bthm} yields the theorem.
\hfill $\Box$

\subsection{Para-fermions of order 2}
For a Young tableau $T$, we denote by $T'$ the tableau obtained by
interchanging the rows and the columns of $T$.
In another word, $T'$ is the transpose of $T$.
The tableau $T_j'$ is on the frame 
$    (\underbrace{2, \cdots, 2}_j, \underbrace{1, \cdots, 1}_{N-2j})      $
and satisfies
\[
       {\cal R}(T'_j) = {\cal C}(T_j), \qquad
         {\cal C}(T'_j)  = {\cal R}(T_j).
\]
The generating functional of the point process for $N$ para-fermions of
order 2 in the finite box $\Lambda_L$ is given by
\[
     E_{L, N}^{2F}\big[e^{-<f, \xi>}\big] =
   \frac{\sum_{j=0}^{[N/2]}\Tr_{\otimes^N\H_L}
            \big((\otimes^N\tilde G) U(d(T'_j))\big)}
        {\sum_{j=0}^{[N/2]}\Tr_{\otimes^N\H_L}
             \big((\otimes^N G) U(d(T'_j))\big)}
\]
as in the case of para-bosons of order 2.
Let us recall the relations
\[
     \chi_{T'_j}(\sigma) = \sgn(\sigma)\chi_{T_j}(\sigma), \qquad
     \varphi_{T'_j}(\sigma) = \sgn(\sigma)\psi_{T_j}(\sigma),
\]
where we have denoted by
\[
     \varphi_{T'_j}(\sigma) = \sum_{\tau}< \tau, \sigma R(b({T'_j}))\tau>
\]
the character of the induced representation 
Ind$_{{\cal C}(T_j')}^{\SN}[ \, \sgn \, ]$, 
where \, sgn\, is the representation 
${\cal C}(T_j') = {\cal R}(T_j) \ni \sigma \mapsto \sgn(\sigma)$.  
Thanks to these relations, we can easily translate the argument of
para-bosons to that of para-fermions and get the following theorem.

\begin{thm}
The finite para-fermion processes defined above converge weakly to the
point process whose Laplace transform is given by
\[
      {\rm E}_{\rho}^{2B}\big[e^{-<f, \xi>}\big]
       = \Det \big[1 - \sqrt{1-e^{-f}}z_*G(1 + z_*G)^{-1}\sqrt{1-e^{-f}}\big]^2
\]
in the thermodynamic limit, where $z_*\in ( 0, \infty)$ is determined by
\[
      \frac{\rho}{2} = \int \frac{dp}{(2\pi)^d}\frac{z_*e^{-\beta |p|^2}}
    {1 + z_*e^{-\beta |p|^2}}= (z_*G(1 + z_*G)^{-1})(x,x).
\]
\end{thm}

\bigskip

\section{Gas of composite particles}
Most gases are composed of composite particles.
In this section, we formulate point processes which yield the position 
distributions of constituents of such gases.
Each composite particle is called a ``molecule",  and molecules consist of 
``atoms".
Suppose that there are two kinds of atoms, say A and B, such that
both of them obey Fermi-Dirac or Bose-Einstein statistics simultaneously,
that $N$ atoms of kind A and $N$ atoms of kind B are in the same box $\Lambda_L$
and that one A-atom and one B-atom are bounded to form a molecule by
the non-relativistic interaction described by the Hamiltonian
\[
        H_L = -\triangle_x -\triangle_y + U(x-y)
\]
with periodic boundary conditions in $L^2( \Lambda_L\times\Lambda_L)$.
Hence there are totally $N$ such molecules in $\Lambda_L$.
We assume that the interaction between atoms in different molecules
 can be neglected.
We only consider such systems of zero temperature, where
 $N$ molecules are  in the ground state and
 (anti-)symmetrizations of the wave functions of the
$N$ atoms of type A and the $N$ atoms of type B are considered.
In order to avoid difficulties due to boundary conditions,
we have set the masses of two atoms A and B equal.
We also assume that the potential $U$ is infinitely deep so that
the wave function of the ground state has a compact support.
We put
\[
    H_L = -\frac{1}{2}\triangle_R - 2\triangle_r + U(r)
    = H_L^{(R)} + H_L^{(r)},
\]
where $R = (x+y)/2, \quad r= x-y $.
The normalized wave function of the ground state of $H_L^{(R)}$ is the
constant
 function $L^{-d/2}$.
Let $\varphi_L(r)$ be that of the ground state of $H_L^{(r)}$.
Then, the ground state of $H_L$ is $\psi_L( x, y) = L^{-d/2}\varphi_L( x -
y)$.
The ground state of the $N$-particle system in $\Lambda_L$ is, by taking the
(anti-)symmetrizations,
\begin{eqnarray}
     \Psi_{L, N}( x_1, \cdots, x_N; y_1, \cdots, y_N)
      &=& Z_{c\alpha}^{-1}\sum_{\sigma, \tau\in\SN}\alpha^{N-\nu(\sigma)}
       \alpha^{N-\nu(\tau)}
          \prod_{j=1}^N\psi_L(x_{\sigma(j)}, y_{\tau(j)}) \nonumber \\
      &=& \frac{N!}{Z_{c\alpha}L^{dN/2}}\sum_{\sigma}\alpha^{N-\nu(\sigma)}
         \prod_{j=1}^N\varphi_L( x_j - y_{\sigma(j)}),
\label{gscs}
\end{eqnarray}
where $Z_{c\alpha}$ is the normalization constant and $\alpha = \pm 1$.
Recall that $\alpha^{N-\nu(\sigma)} = \sgn(\sigma)$ for $\alpha = -1$.

The distribution function of positions of $2N$-atoms of the system
with
zero temperature is given by the square of magnitude of (\ref{gscs})
\begin{equation}
  p^{c\alpha}_{L, N}( x_1, \cdots, x_N; y_1, \cdots, y_N)
  = \frac{(N!)^2}{Z_{c\alpha}^2L^{dN}}\sum_{\sigma, \tau \in\SN}
  \alpha^{N-\nu(\sigma)}\prod_{j=1}^{N}\varphi_L( x_j- y_{\tau(j)})
  \overline{\varphi_L( x_{\sigma(j)}- y_{\tau(j)})}.
\label{epdc}
\end{equation}
Suppose that we are interested in one kind of atoms, say of type A.
We introduce the operator $ \varphi_L $  on $\H_L = L^2(\Lambda_L)$
which has the integral kernel $\varphi_L(x-y)$.
Then the Laplace transform of the distribution of the positions of $N$ A-atoms
 can be written as
\begin{eqnarray*}
  E_{L,N}^{c\alpha}\big[e^{- < f, \xi>}\big]
            &=&\int_{\Lambda^{2N}}e^{-\sum_{j=1}^Nf(x_j)}
            p^{c\alpha}_{L, N}( x_1, \cdots, x_N; y_1, \cdots, y_N)
            \,dx_1\cdots dx_Ndy_1\cdots dy_N \nonumber \\
             &=& \frac{\sum_{\sigma\in\SN}\alpha^{N-\nu(\sigma)}
   \Tr_{\otimes^N\H}[(\otimes^N\varphi_L^*e^{-f}\varphi_L) U(\sigma)]}
        {\sum_{\sigma\in\SN}\alpha^{N-\nu(\sigma)}
       \Tr_{\otimes^N\H}[(\otimes^N \varphi_L^*\varphi_L )U(\sigma)]}.
\end{eqnarray*}

In order to take the thermodynamic limit $ N, L \to \infty, V/L^d \to \rho$,
we consider a Schr\"odinger operator in the whole space.
Let $ \varphi $ be the normalized wave function of the ground state of
$ H_r = - 2\triangle_r + U(r) $ in $L^2(\R^d)$. 
Then
$ \varphi(r) = \varphi_L(r) \; (\forall r \in \Lambda_L)$ holds for
large $L$ by the assumption on $U$.
The Fourier series expansion of $\varphi_L$ is given by
\[
     \varphi_L(r) = \sum_{k\in
\Z^d}\Big(\frac{2\pi}{L}\Big)^{d/2}\hat\varphi
     \Big(\frac{2\pi k}{L}\Big) \frac{e^{i2\pi k\cdot r/L}}{L^{d/2}},
\]
where $\hat\varphi$ is the Fourier transform of $\varphi$:
\[
     \hat\varphi(p) = \int_{\R^d}\varphi(r)e^{-ip\cdot r}
     \frac{dr}{(2\pi)^{d/2}}.
\]
By $\varphi$, we denote the integral operator on $ \H=L^2(\R^d)$ having
kernel $\varphi(x-y)$.

Now we have the following theorem on the thermodynamic limit,
where the density $\rho > 0 $ is arbitrary for $\alpha = -1$, 
$\rho \in ( 0, \rho_c^c)$ for $\alpha =1$ and
\[
        \rho_c^c = \int \frac{dp}{(2\pi)^d}\frac{|\hat\varphi(p)|^2 }
              {|\hat\varphi(0)|^2 - |\hat\varphi(p)|^2 }.
\]
\begin{thm}
The finite point processes defined above for $\alpha = \pm 1$ converge weakly 
to the process whose Laplace transform is given by
\[
      {\rm E}_{\rho}^{c\alpha}\big[e^{-<f, \xi>}\big]
       = \Det \big[1 + z_*\alpha \sqrt{1-e^{-f}}\varphi(||\varphi||_{L^1}^2-z_*\alpha
    \varphi^*\varphi)^{-1}\varphi^*\sqrt{1-e^{-f}}\big]^{-1/\alpha}
\]
in the thermodynamic limit (\ref{tdl}), 
where the parameter $z_*$ is the positive constant uniquely determined by
\[
  \rho = \int \frac{dp}{(2\pi)^d}\frac{z_*|\hat\varphi(p)|^2 }
  {|\hat\varphi(0)|^2 - z_*\alpha|\hat\varphi(p)|^2 } =
     (z_*\varphi(||\varphi||_{L^1}^2-z_*\alpha\varphi^*\varphi)^{-1}\varphi^*)(x,x).
\]
\label{cthm}
\end{thm}
\noindent{\sl Proof : } The eigenvalues of the integral operator 
$\varphi_L$ is $\{(2\pi)^{d/2}\hat\varphi(2\pi k/L)\}_{k\in\Z^d}$. 
Since $\varphi$ is the ground state of the Schr\"odinger operator, 
we can assume $\varphi \geqslant 0$.
Hence the largest eigenvalue is $(2\pi)^{d/2}\hat\varphi(0)= ||\varphi||_{L^1}$.
We also have 
\begin{equation}
     1 = ||\varphi||^2_{L^2(\R^d)} = \int_{\R^d}|\hat \varphi(p)|^2dp
     = ||\varphi_L||^2_{L^2(\Lambda_L)}
     = \sum_{k\in \Z^d}\Big(\frac{2\pi}{L}\Big)^d
      \Big|\hat\varphi\Big(\frac{2\pi k}{L}\Big)\Big|^2.
\label{s=s}
\end{equation}
Set $ V_L=\varphi_L/||\varphi||_{L^1}$ so thet
\[
  ||V_L|| = 1, \quad ||V_L||_2^2 = L^d/||\varphi||_{L^1}^2.
\]
Then Theorem \ref{gthm} applies as follows:

For $z \in I_{\alpha}$, let us define functions $d, d_L$ on $\R^d$ by
\[
    d(p) = \frac{z|\hat\varphi(p)|^2}
    {|\hat\varphi(0)|^2 - z\alpha|\hat\varphi(p)|^2}
\]
and
\begin{equation}
     d_L(p) =d(2\pi k/L) \quad\mbox{if} \quad
     p\in \Box_k^{(L)}
     \quad \mbox{for} \quad k \in \Z^d.
\label{d_L}
\end{equation}
Then
\[
   \int_{\R^d}\frac{dp}{(2\pi)^d}d_L(p)
   = L^{-d}||zV_L(1 - z\alpha V_L^*V_L)^{-1}V_L^*||_1
\]
and the following lemma holds:
\begin{lem}
\[
    \lim_{L\to\infty}|| d_L - d ||_{L^1} = 0.
\]
\end{lem}
{\sl Proof : } Put
\[
     \hat\varphi_{[L]}(p) =\hat\varphi(2\pi k/L) \quad\mbox{if} \quad
     p\in \Box_k^{(L)}
     \quad \mbox{for} \quad k \in \Z^d
\]
and note that compactness of supp$\;\varphi$ implies 
 $\varphi\in L^1(\R^d)$ and uniform continuity of $\hat\varphi$.
Then we have 
$ ||\,|\hat \varphi_{[L]}|^2 - |\hat \varphi|^2 ||_{L^{\infty}} \to 0 $
and $ || d_L - d ||_{L^{\infty}} \to 0 $.
On the other hand, we get 
$ ||\,|\hat \varphi_{[L]}|^2||_{L^1} = ||\, |\hat \varphi|^2 ||_{L^1} $
from (\ref{s=s}).
It is obvious that
\[
     |\, || d_L ||_{L^1}- ||d ||_{L^1} \,|\leqslant
             \frac{z}{(1-z(\alpha\vee 0))^2}
    \frac{||\,|\hat \varphi_{[L]}|^2 - |\hat \varphi|^2 ||_{L^1}}
     {|\varphi(0)|^2}.
\]
Hence the lemma is derived by using the following fact twice:

If $ f, f_1, f_2, \cdots \in L^1(\R^d)$ satisfy
\[
     ||f_n - f||_{L^{\infty}} \to 0 \; \mbox{ and }
     \; ||\, f_n\,||_{L^1} \to ||\, f\, ||_{L^1},
\]
then   $ ||f_n - f||_{L^1} \to 0 $ holds.

In fact, using
\[
     \int_{|x| > R}|f_n(x)|\,dx = \int_{|x| > R}|f(x)|\,dx
     + \int_{|x| \leqslant R}(|f(x)| - |f_n(x)|)\,dx +
     ||\, f_n\, ||_{L^1} - ||\, f\, ||_{L^1},
\]
we have
\begin{eqnarray*}
|| f_n - f||_{L^1} 
     &\leqslant & \int_{|x| \leqslant R}|f_n(x) - f(x)|\,dx
       + \int_{|x| > R}(|f_n(x)| + |f(x)|)\,dx  \\
&\leqslant &  2\int_{|x| \leqslant R}|f_n(x) - f(x)|\,dx
    + 2\int_{|x| > R} |f(x)|\,dx + ||\, f_n\, ||_{L^1} - ||\, f\, ||_{L^1}.
\end{eqnarray*}
For any $\epsilon >0$, we can choose $R$ large enough to make the second
 term of the right hand side smaller than $\epsilon$.
For this choice of $R$, we set $n$ so large that the first term and the
remainder are smaller than $\epsilon$ and then $|| f_n - f ||_{L^1} < 3\epsilon$.  
\hfill $\Box$

\bigskip
\noindent ( {\sl Continuation of the proof of Theorem \ref{cthm}} ) 
Using this lemma, we can show
\begin{eqnarray*}
&&  h_L^{(\alpha)}(z) = 
 \frac{\Tr [zV_L(1-z\alpha V_L^*V_L)^{-1}V_L^*]}{\Tr V_L^*V_L}
 \quad \to \quad |\hat\varphi(0)|^2\int_{\R^d}dp
       \frac{z|\hat\varphi(p)|^2}{|\hat\varphi(0)|^2-z\alpha |\hat\varphi(p)|^2} 
       = h^{(\alpha)}(z),
\\
&& ||\sqrt{1-e^{-f}}\big[V_L(1-z\alpha V_L^*V_L)^{-1}V_L^*
   - \varphi(||\varphi||_{L^1}^2-z\alpha \varphi^*\varphi)^{-1}\varphi\big]
   \sqrt{1-e^{-f}}||_1\to 0, 
\end{eqnarray*}
as in the proof of (\ref{falpha}) and (\ref{limK}). 
We have the conversion $\hat\rho = ||\varphi||_{L^1}^2\rho$
and hence $\rho_c^c= \sup_{z\in I_1}h^{(1)}(z)/||\varphi||_{L^1}^2$.
Hence the proof is completed by Theorem \ref{gthm}.
\hfill   $\Box$

\bigskip

\bigskip

\noindent{\bf\Large Acknowledgements}

\bigskip

\noindent We would like to thank Professor Y. Takahashi and Professor T. Shirai 
for many useful discussions.
K. R. I. thanks Grant-in-Aid for Science Research (C)15540222 from JSPS.

\bigskip

\appendix
\section{ Complex integrals }
\begin{lem}
   {\rm (i)}  For $ 0 \leqslant p \leqslant 1 $ and $ -\pi \leqslant \theta \leqslant \pi $,
\[
         |1+ p(e^{i\theta}-1)| \leqslant \exp
         \Big(-\frac{2p(1-p)\theta^2}{\pi^2}\Big)
\]
holds.  For $ 0 \leqslant p \leqslant 1 $ and $ -\pi/3 \leqslant \theta \leqslant \pi/3 $,
\[
       |\log \big(1+p(e^{i\theta}-1)\big)
       - ip\theta + \frac{p(1-p)}{2}\theta^2| \leqslant
        \frac{4p(1-p)|\theta|^3}{9\sqrt 3}
\]
holds.

{\rm (ii)} For $ p \geqslant 0 $ and $ -\pi \leqslant \theta \leqslant \pi$,
 the following inequalities hold.
\begin{eqnarray*}
        |1-p(e^{i\theta}-1)| 
        &\geqslant&  \exp\Big(\frac{2p(1+p)}{1+4p(1+p)}
               \frac{\theta^2}{\pi^2}\Big) \\
       |\log \big(1-p(e^{i\theta}-1)\big)
              +ip\theta - \frac{p(1+p)}{2}\theta^2| 
        &\leqslant&       \frac{p(1+p)(1+2p)|\theta|^3}{6}
\end{eqnarray*}
\end{lem}
{\sl Proof: }  (i) The first inequality follows from

\begin{equation}
    |1+ p(e^{i\theta}-1)|^2 = 1 - 2p(1 - p)(1-\cos \theta)
\label{+p}
\end{equation}
\[
    \leqslant\exp(-2p(1-p)(1-\cos \theta))
     \leqslant \exp(-4p(1-p)\theta^2/\pi^2),
\]
where $ 1- \cos\theta \geqslant 2\theta^2/\pi^2 $
for $ \theta \in [-\pi, \pi] $ is used in the second inequality.

Put $ f(\theta) = \log(1+ p(e^{i\theta}-1)) $. 
Then we have $f(0)=0$,
\begin{eqnarray*}
   f'(\theta) &=& i - \frac{i(1-p)}{1-p+pe^{i\theta}}, \quad f'(0) = ip,
             \\    f''(\theta) &=&
  - \frac{p(1-p)e^{i\theta}}{(1-p+pe^{i\theta})^2},
      \quad f''(0) = -p(1-p)
\end{eqnarray*}
and
\[
    f^{(3)}(\theta) = - \frac{ip(1-p)e^{i\theta}(1-p-pe^{i\theta})}
       {(1-p+pe^{i\theta})^3}.
\]
By (\ref{+p}) and $ \theta\in[-\pi/3, \pi/3]$, we have
$|1+ p(e^{i\theta}-1)|^2 \geqslant 1-p(1-p)\geqslant 3/4$.
Hence, $ |f^{(3)}(\theta)| \leqslant 8p(1-p)/3\sqrt 3$ holds.
Taylor's theorem yields the second inequality.

\medskip
(ii)  The first inequality follows from
\begin{eqnarray*}
\lefteqn{ |1-p(e^{i\theta}-1)|^2 = 1 + 2p(1+p)(1-\cos \theta) 
        } && \\
     &\geqslant& 
     \exp\Big(\frac{2p(1+p)(1-\cos \theta)}{1+4p(1+p)}\Big)
     \geqslant \exp\Big(\frac{4p(1+p)}{1+4p(1+p)}
     \frac{\theta^2}{\pi^2}\Big).
\end{eqnarray*}
     Here we have used $1+x \geqslant e^{x/(1+a)} $ for $ x\in [0, a] $
in the first inequality, which is derived from
$ \log(1+x) = \int_0^xdt/(1+t) \geqslant x/(1+a)  $.

Put $ f(\theta) = \log(1-p(e^{i\theta}-1)) $. Then we have $f(0)=0$,
\begin{eqnarray*}
   f'(\theta) &=&
        i - \frac{i(1+p)}{1+p-pe^{i\theta}}, \quad f'(0) =-ip, \\
   f''(\theta) &=&
       \frac{p(1+p)e^{i\theta}}{(1+p-pe^{i\theta})^2},
      \quad f''(0) = p(1+p)
\end{eqnarray*}
and
\[
    f^{(3)}(\theta) = \frac{ip(1+p)e^{i\theta}(1+p+pe^{i\theta})}
       {(1+p-pe^{i\theta})^3}.
\]
Hence, we have $ |f^{(3)}(\theta)| \leqslant p(1+p)(1+2p)$.
Thus we get the second inequality.
\hfill $\Box$

\begin{prop}
Let $ s > 0$ and a collection of numbers $\{\, p^{(N)}_j \, \}_{j,N}$
satisfy
\[
   p^{(N)}_0 \geqslant  p^{(N)}_1 \geqslant p^{(N)}_2 \geqslant \cdots
   \geqslant p^{(N)}_j \geqslant \cdots \geqslant 0, \quad
   \sum_{j=0}^{\infty}sp^{(N)}_j = N.
\]

\medskip
\noindent {\rm (i)} Moreover, if $\, p^{(N)}_0 \leqslant 1 \,$ and
\[
   v^{(N)} \equiv \sum_{j=0}^{\infty}
   sp^{(N)}_j ( 1 - p^{(N)}_j) \rightarrow \infty \quad ( N \to \infty ),
\]
then
\[
         \lim_{N\to\infty}\sqrt{v^{(N)}} \oint_{S_1(0)}\frac{d\eta}{2\pi i
         \eta^{N+1}} \prod_{j=0}^{\infty}(1+p_j^{(N)}(\eta-1))^s
                  =  \frac{1}{\sqrt{2\pi}}
\]
holds.

\medskip
\noindent {\rm (ii)} If $\, \{p^{(N)}_0 \}$ is  bounded,  
then
\[
     \lim_{N\to\infty}\ \sqrt{w^{(N)}} \oint_{S_1(0)}\frac{d\eta}{2\pi i\eta^{N+1}}
     \frac{1}{\prod_{j=0}^{\infty}(1 - p_j^{(N)}
         (\eta-1))^s} =  \frac{1}{\sqrt{2\pi}}
\]
holds, where 
\[
    w^{(N)} \equiv \sum_{j=0}^{\infty}sp^{(N)}_j ( 1 + p^{(N)}_j).
\]
\end{prop}
{\sl Proof:} 
\noindent (i) Set $ \eta = \exp(ix/\sqrt{v^{(N)}})$. 
Then the integral is written as $ \int_{-\infty}^{\infty}h_N(x)\,dx/2\pi$,
where
\[
        h_N(x) = \chi_{[-\pi\sqrt {v^{(N)}}, \pi\sqrt{v^{(N)}}]}(x)
            e^{-iNx/\sqrt {v^{(N)}}}
            \prod_{j=0}^{\infty}\big[1+p_j^{(N)}
          (e^{ix/\sqrt {v^{(N)}}} -1)\big]^s.
\]
By Lemma A.1(i), we have
\[
    |h_N(x)| \leqslant \prod_{j=0}^{\infty}
      e^{-2sp_j^{(N)}(1-p_j^{(N)})x^2/\pi^2 v^{(N)}} = e^{-2x^2/\pi^2}
         \in L^1(\R).
\]
If $N$ is so large that $ |x/\sqrt{v^{(N)}}|\leqslant \pi/3 $, we also get
\begin{eqnarray*}
  h_N(x)&=& \chi_{[-\pi\sqrt {v^{(N)}}, \pi\sqrt {v^{(N)}}]}(x)
           \exp\Big[-i\frac{Nx}{\sqrt{v^{(N)}}} + s\sum_{j=0}^{\infty}
           \log\big(1+ p_j^{(N)}
          (e^{ix/\sqrt {v^{(N)}}} -1)\big)\Big] \\
        &=& \chi_{[-\pi\sqrt {v^{(N)}}, \pi\sqrt {v^{(N)}}]}(x)
           \exp\Big[-i\frac{Nx}{\sqrt{v^{(N)}}} + s\sum_{j=0}^{\infty}
           \big( i\frac{p_j^{(N)}x}{\sqrt{v^{(N)}}}
          - \frac{p_j^{(N)}(1-p_j^{(N)})x^2}{2v^{(N)}} +
              \delta^{(N)}_j\big)\Big] \\
        &=& \chi_{[-\pi\sqrt {v^{(N)}}, \pi\sqrt {v^{(N)}}]}(x)
             \exp\big(-\frac{x^2}{2}+ \delta^{(N)}\big)
           \underset{N\to\infty}{\longrightarrow} e^{-x^2/2},
\end{eqnarray*}
where
\[
             |\delta^{(N)}| = |\sum_{j=0}^{\infty}s\delta^{(N)}_j| \leqslant
            \sum_{j=0}^{\infty} \frac{4sp_j^{(N)}(1-p_j^{(N)})x^3}
            {9\sqrt 3\sqrt{v^{(N)}}^3}
             = \frac{4|x^3|}{9\sqrt {3v^{(N)}}}.
\]
The dominated convergence theorem yields
\[
        \int_{-\infty}^{\infty}h_N(x)\frac{dx}{2\pi}
\underset{N\to\infty}{\longrightarrow}
        \int_{-\infty}^{\infty}\frac{dx}{2\pi} e^{-x^2/2} =
\frac{1}{\sqrt{2\pi}}.
\]

\medskip
\noindent (ii) Note that $w^{(N)} \to \infty$ as $N \to \infty$.
Set $ \eta = \exp(ix/\sqrt {w^{(N)}})$. 
Then the integral is written as $ \int_{-\infty}^{\infty}k_N(x)\,dx/2\pi$,
where
\[
        h_N(x) = \chi_{[-\pi\sqrt {w^{(N)}}, \pi\sqrt{w^{(N)}}]}(x)
            \frac{e^{-iNx/\sqrt {w^{(N)}}}}
            {\prod_{j=0}^{\infty}\big[1-p_j^{(N)}
          (e^{ix/\sqrt {w^{(N)}}} -1)\big]^s}.
\]
By Lemma A.1(ii) and the boundedness of $\{p_0^{(N)}\}$,
we have, with some positive constant $c$,
\[
    |k_N(x)| \leqslant \prod_{j=0}^{\infty}
      \exp\Big(-\frac{2sp_j^{(N)}(1+p_j^{(N)})}{1+4p_0^{(N)}(1+p_0^{(N)})}
     \frac{x^2}{\pi^2 w^{(N)}}\Big) \leqslant  e^{-cx^2}
        \in  L^1(\R)
\]
and
\begin{eqnarray*}
 k_N(x) &=& \chi_{[-\pi\sqrt {w^{(N)}}, \pi\sqrt {w^{(N)}}]}(x)
           \exp\Big[-i\frac{Nx}{\sqrt{w^{(N)}}} - s\sum_{j=0}^{\infty}
           \log\big(1- p_j^{(N)}
          (e^{-ix/\sqrt {w^{(N)}}} -1)\big)\Big] \\
        & =&  \chi_{[-\pi\sqrt {w^{(N)}}, \pi\sqrt {w^{(N)}}]}(x)
           \exp\Big[-i\frac{Nx}{\sqrt{w^{(N)}}} - s\sum_{j=0}^{\infty}
           \big( -i\frac{p_j^{(N)}x}{\sqrt{w^{(N)}}}
        + \frac{p_j^{(N)}(1+p_j^{(N)})x^2}{2w^{(N)}} +
              \delta^{(N)}_j\big)\Big] \\
        &=&  \chi_{[-\pi\sqrt {w^{(N)}}, \pi\sqrt {w^{(N)}}]}(x)
             \exp\big(-\frac{x^2}{2}+ \delta^{(N)}\big)
\underset{N\to\infty}{\longrightarrow} e^{-x^2/2},
\end{eqnarray*}
where
\[
             |\delta^{(N)}| = |\sum_{j=0}^{\infty}s\delta^{(N)}_j| \leqslant
            \sum_{j=0}^{\infty} 
            \frac{p_j^{(N)}(1+p_j^{(N)})(1+ 2p^{(N)}_j)|x^3|}
            {6\sqrt{w^{(N)}}^3} \leqslant 
            \frac{(1+2p_0^{(N)})}{6\sqrt{w^{(N)}}}|x^3|.
\]
The result is obtained by the dominated convergence theorem. \hfill $\Box$

\end{document}